%% file: paper-Transients-SGRA.tex
\documentclass[traditabstract,twocolumn]{aa}

\usepackage{epsfig,graphicx,natbib}
\usepackage{longtable}
\usepackage{amssymb}
\usepackage{amsmath}
\usepackage{color}
\usepackage{subcaption}
\usepackage{booktabs}

\usepackage{xspace}
\usepackage{float}
\usepackage{longtable}
\usepackage{breqn}
\usepackage{adjustbox}
\usepackage{booktabs,caption}
\usepackage[flushleft]{threeparttable}

\usepackage{todonotes}

\newcommand{\deltara}{\Delta \text{ra}}
\newcommand{\deltadec}{\Delta \text{dec}}
\newcommand{\sgra}{Sgr~A$^*$\xspace}

\begin{document}

\newcommand{\fmini}{F^{\text{ext}}}
\newcommand{\fsgra}{F^{\text{SgrA*}}}
\newcommand{\vmod}{\mathcal{V}_{t}^{\text{ext}}}
\newcommand{\vmmod}{\mathcal{V}_{t}^{\text{mod}}} 
\newcommand{\vobs}{\mathcal{V}_{t,k}^{\text{obs}}}
\newcommand{\vobsnb}{\mathcal{V}_{t}^{\text{obs}}}

\title{A first search of transients in the Galactic Center from 230 GHz ALMA observations}

\author{A.~Mus \inst{1,2}
\and I. Mart\'i-Vidal \inst{1,2}
\and M. Wielgus \inst{3}
\and G. Stroud \inst{4}
}

\institute{
  Departament d’Astronomia i Astrof\'isica, Universitat de Val\`encia, C. Dr. Moliner 50, 46100 Burjassot ,Val\`encia, Spain
  \and
  Observatori Astron\`omic, Universitat de Val\`encia, Parc Cient\'ific, C. Catedr\`atico Jos\'e Beltr\'an 2, 46980 Paterna, Val\`encia, Spain
  \and
  Max-Planck-Institut f\"ur Radioastronomie, Auf dem H\"ugel 69, D-53121 Bonn, Germany
  \and
  Department of Physics, Imperial College London, London, United Kingdom
}

\date {Received  / Accepted}

\titlerunning{A search for radio transients in the Galactic Center using ALMA}
\authorrunning{Mus A, Mart\'i-Vidal I et al.}

\abstract{
The Galactic Center (GC) presents one of the highest stellar densities in our Galaxy, making its surroundings an environment potentially rich in radio transients, such as pulsars and different kinds of flaring activity. In this paper, we present the first study of transient activity in the region of the GC based on Atacama Large Millimeter/submillimeter (mm/submm) Array (ALMA) continuum observations at 230 GHz. This search is based on a new self-calibration algorithm, especially designed for variability detection in the GC field.
Using this method, we have performed a search of radio transients in the effective field of view of~$\sim 30\,$arcseconds of the GC central supermassive black hole Sagittarius A* (\sgra) using ALMA 230 GHz observations taken during the 2017 Event Horizon Telescope (EHT) campaign, which span several observing hours (5-10) on 2017 April 6, 7, and 11.
This calibration method allows one to disentangle the variability of unresolved \sgra from any potential transient emission in the wider field of view and residual effects of the imperfect data calibration. Hence, a robust statistical criterion to identify real transients can be established: the event should survive at least three times the correlation time and it must have a peak excursion of at least seven times the instantaneous root-mean-square between consecutive images.
Our algorithms are successfully tested against realistic synthetic simulations of transient sources in the GC field. Having checked the validity of the statistical criterion, we provide upper limits for transient activity in the effective field of view of the GC at 230 GHz.
}

\maketitle

\section{Introduction}

Many astronomical compact radio sources manifest as transients~\citep[e.g.,][]{Cordes04,Lazio08,Lorimer18,Murphy21}. Usually, such sources originate from explosive or dynamic events, for example, 
outbursting X-ray binaries~\citep{Bower05,Generozov18}. Transient radiation emission is hence a signature of some of the most interesting physical phenomena in the Universe. In particular, one of the most studied regions where transients can be found is in the Galactic Center (GC). The time variability of the physical conditions that happen in the central supermassive black hole of the GC Sagittarius A* (\sgra) is particularly interesting since its astrophysical interactions with its immediate surroundings hints at potential transient radiation on timescales of seconds to minutes.
Infrared observations revealed the presence of a dense population of young stars in the GC region, commonly known as the S-star cluster \citep{Eckart96, Ghez98, Schodel02, Eisenhauer05}. Near-infrared and X-ray observations have detected flares occurring around \sgra~\citep{Genzel03, Aschenbach04,Dolence12}. Diverse motivations exist to search for transients in the GC region; for example, if we focus on the immediate vicinity of the central black hole, \sgra, their detection on short timescales could constrain the population of pulsars that exist in that region~\citep{Kuo21} as well as shed light on the accretion of \sgra~\citep{Eatough13, Desvignes18}, or even measure, with a never-achieved precision, the mass and spin of \sgra~\citep{Wex99, Kramer04, Liu12, Psaltis16, Liu17, Laurentis18}.
The GC field at parsec scales is surrounded by extended thermal structures called ``minispiral,'' which are likely related to dust and ionized gas~\citep[e.g.,][]{Lo83,Mills17,Tsuboi18,Bhat22}.
There have been several searches for transients on timescales of days to years in the inner parsec of \sgra ~\citep{Hyman05, Bower07, Bower11, Bell11, Thyagarajan11, Mooley13, Chiti16, Kuo21, Eatough21}. In particular, the most remarkable pulsars are the six radio-emitting neutron starts within half a degree from \sgra~\citep{Johnston06, Deneva09, Eatough13,Zhao22} and magnetar PSR J1745-2900~\citep{Mori13, Rea13,Esposito21}, which is inhabiting the innermost parsec region. Nevertheless, the detection of transient sources is well below the expected number of pulsars in the GC region~\citep[e.g.,][]{Chennamangalam14, Zhang14}. This could be due to the challenges of detecting these transient sources~\citep{Cordes03}, compared to persistent ones.  Since the transient emission can be visible only within a short timescale, radio transients are often associated with coherent emission processes~\citep[e.g.,][]{Hyman05,Wang21}, in addition to powerful probes of intervening media, for instance, the scattering effect seen along the line of sight to the GC may broaden intrinsically narrow pulses. 

Observations of transient emission at short -- millimeter (mm) -- wavelengths are challenging from a technical standpoint. On the one hand, source-intrinsic effects (i.e., spectra with lower intensities at higher frequencies as pulsars) limit the observability of relatively weaker transients. On the other hand, instrumental effects (lower sensitivities and unstable electronics) and atmospheric issues (higher opacity and phase variability) limit the fidelity of variability detection.
Very recently, \cite{Kuo21} have reported an ALMA transient and pulsar search in the GC at 86 GHz, based on the Global Millimiter VLBI Array campaign the 2017 April 3 and 2018 April 14, April 17 at 3.5\,mm. These authors do not report any transient detection at that wavelength.

In this work, we present a novel and nonstandard self-calibration procedure that overcomes most of the instrumental and atmospheric limitations of mm-wave transient studies. Our method is based on the correction of instrumental and atmospheric amplitude variations by using an intra-source visibility scaling tied to the extended (hence stable in time) mini-spiral emission.
We then apply this method to the ALMA observations of \sgra at 230 GHz taken during the EHT 2017 campaign~\citep{EHT1,EHT2,EHT3,EHT4,EHT5,EHT6,CiriacoRef} (hereafter Paper I to Paper IV respectively) and perform a transient search in the GC region at parsec scales. Thanks to the superb sensitivity of ALMA, the transient search reported here is the most sensitive one (1.0\,mJy/beam in terms of flux density per unit time) reported to date at mm wavelengths. Indeed, and as we discuss in Sect.~\ref{sec:realdata}, we are limited by dynamic range (rather than by sensitivity) even at the level of the integration time used in the ALMA correlator (four seconds).
\\
In Sect.~\ref{sec:observations}, we describe the details of the observations reported in this work. In Sect.~\ref{sec:selfcal_approach}, we describe the novel self-calibration procedure in full detail. In Sect~\ref{sec:stgy} we explain the statistics and the criterion we have developed to consider an event as a transient. In Sect.~\ref{sec:synthetic_data}, we demonstrate the effectiveness of our new technique by searching for a transient included in a synthetic dataset. Finally, in Sect.~\ref{sec:realdata}, we report on the results of our transient detection method, as applied to the real ALMA observations.

\section{Observations}
\label{sec:observations}
The ALMA observatory is able to combine the signal of all its antennas into a ``phased sum'', suited for its correlation as a single very-long-baseline interferometry (VLBI) station~\citep[e.g.][]{APPRef}. ALMA is a key station participating in the EHT observing campaigns since 2017, offering a huge sensitivity boost~(Paper I to Paper III) and serving as reference station for coherent fringe search algorithms~\citep{Blackburn19,Janssen19}.
The VLBI observations of ALMA during year 2017 were part of its Cycle 4 campaign, which was carried out in compact configuration C40-3 (baseline lengths between 15\,m and 460\,m). The campaign was divided in two parts, each one observing in a different frequency band: April 1 to April 3 in Band 3 (86\,GHz) and April 5 to April 11 in Band 6 (230\,GHz). In this work, we focus on 230\,GHz. For the frequency set up there were two sidebands (called ``lower sideband" and ``upper sideband", or LO and HI) with four frequency spectral windows (two per sideband), 
with a bandwidth of 2\,GHz each (divided in 240 spectral channels) two centered at 213.1 and 215.1\,GHz (lower sideband), and two centered at 227.1 and 229.1\,GHz (upper sideband). The intra-ALMA visibilities were a ``byproduct" of the VLBI observations.

The GC was observed in three tracks, hereafter called Track B (April 6), Track C (April 7), and Track E (April 11)  with a typical duration of 4-10 hours and a median number of 37 antennas of 12\,m diameter giving a effective field of view of the GC field of~$\sim30$\,arcsec. The integration time of the ALMA correlator in VLBI mode is set to 4 seconds, which corresponds to the shortest cadence for our variability study, resulting in a sensitivity, in standard weather conditions, of 0.5\,mJy/beam to 1.0\,mJy/beam. In our observations we get greater values, which is a hint that we are dynamic range limited. More details can be found in~\cite{CiriacoRefb} and~\cite{Wielgus21}.

As described in~\cite{APPRef} and~\cite{CiriacoRefb}, the ALMA phased-array operations require a nonstandard calibration procedure, since the data path from the antennas to the ALMA correlator differs from the standard interferometric operations and some particular hardware corrections are disabled in VLBI mode. For this work, we have applied the visibility calibration following the procedure of~\cite{CiriacoRefMini}  plus subsequent calibration steps described in~\cite{Wielgus21}.

\section{Two-components' self-calibration}
\label{sec:selfcal_approach}

Standard self-calibration method~\citep{selfcalRef} is a widely extended and commonly used strategy for imaging radio interferometric data. Its goal is to adjust the complex gains of an interferometer, by comparing the calibrated visibilities to an (iteratively improved) model. An accurate modeling of the source structure is crucial for the correct convergence of self-calibration~\citep[e.g.][]{imv2008}. Finding an accurate model of complex structures can be  a challenging task. In the case of the GC (with a rapidly varying compact component, combined with a rich extended mini-spiral structure), finding an appropriate self-calibration model is especially difficult. 

The QA2 calibration ``pipeline"~\citep{APPRef, CiriacoRefb} is the procedure used to generate the intra-ALMA calibration tables and all the metadata needed for the correct polarimetry processing of the ALMA-VLBI observations. This procedure assumes that the observed sources maintain a constant total flux density throughout the observing epoch. However, the high variability of \sgra makes it necessary to refine the QA2 calibration by removing such a restriction. A typical strategy to deal with interferometric observations of a variable compact source is to use only a selection of the longest available baselines~\citep[e.g.][]{Brinkerink15, Iwata20,Wielgus21}. This way, possible artifacts related to the interferometric sampling of the Fourier transform 
of any extended structure is expected to be separated from the time-variable signal of a compact source. This procedure, even though it can reduce the impact of the extended component in the analysis of the \sgra variability, involves the discarding of a large amount of data, which may considerably affect the sensitivity.

In the following section, we present a nonstandard self-calibration procedure, similar to the one described in~\cite{Wielgus21} that allows us, on the one hand, to correct the instrumental and atmospheric effects in the ALMA observations of the GC and, on the other hand, to use the complete set of visibilities (i.e., both long and short baselines) in a self-consistent, precise and sensitive transient analysis.

\begin{figure*}
    \centering
    \hspace{-1cm}
    \includegraphics[scale=0.4]{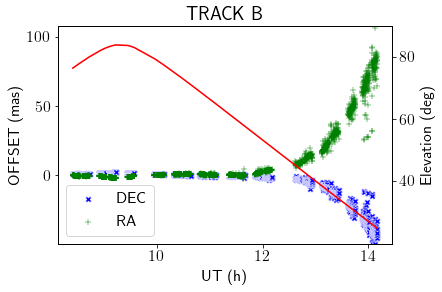}
    \hspace{-0.2cm}
    \includegraphics[scale=0.4]{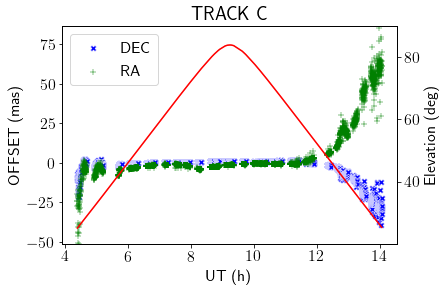}
    \hspace{-0.2cm}
    \includegraphics[scale=0.4]{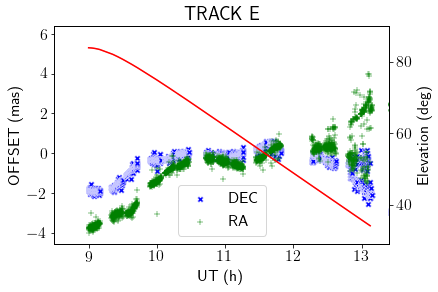}
     
    \caption{\sgra phase-center offsets in right ascension and declination with respect to phase center for each day: Track B (April 6), Track C (April 7) and Track E (April 11). In red, antenna elevation during the tracks.}
    \label{fig:fitting_error}
\end{figure*}

\subsection*{Two-components' self-calibration algorithm}
At the arcsecond scales seen by ALMA, \sgra field of view (FOV) can be decomposed in two parts~\citep[e.g.,][]{CiriacoRefMini,Wielgus21}: an unresolved point source dominated by the compact emission on a scale of at most several Schwarzschild radii from the \sgra supermassive black hole~\citep{Doeleman18} and an extended region called ``minispiral"~\citep[e.g.][]{Lo83}. The compact component has the property of being highly variable at millimeter wavelengths on short timescales, with flux density varying within $3.2 \pm 1.2$\,Jy~\citep[see][]{Wielgus21} The morphology of the large scale minispiral emission, on the other hand, is stable in time on relevant timescales of several days, hence we assume this in the presented analysis. Nevertheless we allow for the total flux density scale of the extended emission component to vary between the three observed tracks, finding values between 1.16 and 1.18 Jy. Wielgus et al. (2022) considered the integrated flux density of the mini-spiral to be $\sim$ 1.1 Jy.

Following~\cite{Wielgus21}, we have modeled these two components across the FOV at each integration time of 4s. In consequence, we define the visibility model, $\mathcal{V}_{j,t}$ for a given baseline, $j$ at a given observing time $t$, as
\begin{equation}\label{eq:model}
\mathcal{V}^{\text{mod}}_{j,t} =\left(\mathcal{V}^{\text{ext}} \fmini_{j,t} + \fsgra\right)\\e^{2\pi i\left(u\deltara + v\deltadec\right)},
\end{equation}

where $u$ and $v$ are the coordinates of the projected baseline in the UV plane at time $t$, $\mathcal{V}^{\mathrm{ext}}$ is the normalized model visibility corresponding to the extended mini-spiral structure (it depends on $u$ and $v$, which in turn depend on the baseline and time), $\fmini_{j,t}$ is the deconvolved flux density of the extended component, which (once subtracted the contribution from \sgra) should correspond to the extrapolated zero-space flux density of the observations, while $\fsgra$ is the total flux density of the central compact source, that is \sgra. The two extra parameters, $\Delta$ra and $\Delta$dec, are time-dependent right ascension and declination offsets of the entire source structure. Such offsets may originate from an imperfect phase calibration, which was originally performed by neglecting the effects of the minispiral at each integration time~\citep{CiriacoRefMini}. Fig.~\ref{fig:fitting_error} shows how optimal elevations present less offset on the relative right ascension (RA) and relative declination (DEC) with respect to \sgra correlation coordinates (17:45:40.04 -29.00.28.17(J2000). That means, the gains for the suboptimal elevations are more affected by atmospheric conditions and more corrections with respect to the QA2 are needed.

The observed visibility vector at a time $t$ given a baseline $j$, denoted by $\mathcal{V}_{j,t}^{\text{obs}}$, can be model-fitted to $\mathcal{V}_{j,t}^{\text{mod}}$ by minimizing the chi-square distribution function
\begin{equation}
\chi^2_t(\fsgra,\fmini_{j,t}) = \sum_j{ \omega_{j,t} \left| \mathcal{V}^{\rm obs}_{j,t} - \mathcal{V}^{\rm mod}_{j,t} \right|^2}, 
\label{Chi2Minispiral:eq}
\end{equation}
where $\mathcal{V}^{\text{mod}}_{j,t}$ is related to $\fsgra,\fmini_{t}$ via Eq.~\ref{eq:model}, $\omega_{j,t}$ is the signal-to-noise baseline weights and the summation extending over all baselines available, $j$, at a given time $t$.

We define the gain  $G_t$ at instant time $t$, as the ratio between the 
fitted flux density of the mini-spiral at time $t$, ($\fmini_t$), and the mean mini-spiral flux density, a real positive number ($F^{\text{mean}}$), i.e.,
\begin{equation*}
    G_t = \dfrac{\fmini_{t}}{F^{\text{mean}}}.
\end{equation*}
Given this definition of $G_t$, the following relation is satisfied for all times $t$:
\begin{equation*}
    \dfrac{\fmini_{t}}{G_t} = F^{\text{mean}}. 
\end{equation*}
Since the integrated flux density of the mini-spiral is assumed to be constant, we can use the values of $G_t$ to remove the residual corruption effects in the \sgra flux density estimates $\fsgra$. Hence, we produce a corrected estimate of the \sgra flux
density, $F^{\text{c}}_t$, using the equation
\begin{equation}
F^{\text{c}}_t = \frac{\fsgra}{G_t}.
\label{CorrFlux:eq}
\end{equation} 

Observe that this procedure, exactly as is done in~\cite{Wielgus21},  would allow us to estimate the time-variable flux density of \sgra using all the available ALMA baselines (i.e., not only the long ones, where the mini-spiral are resolved out), hence maximizing the sensitivity of our analysis.
If the assumption of the constancy of the total mini-spiral’s flux density holds, $G_t$ account for atmospheric effects, meaning that any atmospheric opacity or any loss of phase coherence should be modeled out by the setting of a constant total mini-spiral’s flux density.

In addition to the use of $G_t$, we have also performed one standard self-calibration iteration(i.e., computed the antenna-wise time-dependent complex gains) correcting at the same time amplitudes and phases using the model visibilities from Eq.~\ref{eq:model} (i.e., accounting for the time-variable \sgra plus the mini-spiral), to improve even more the data accuracy.

\begin{figure}
    \centering
    \includegraphics[scale=0.4]{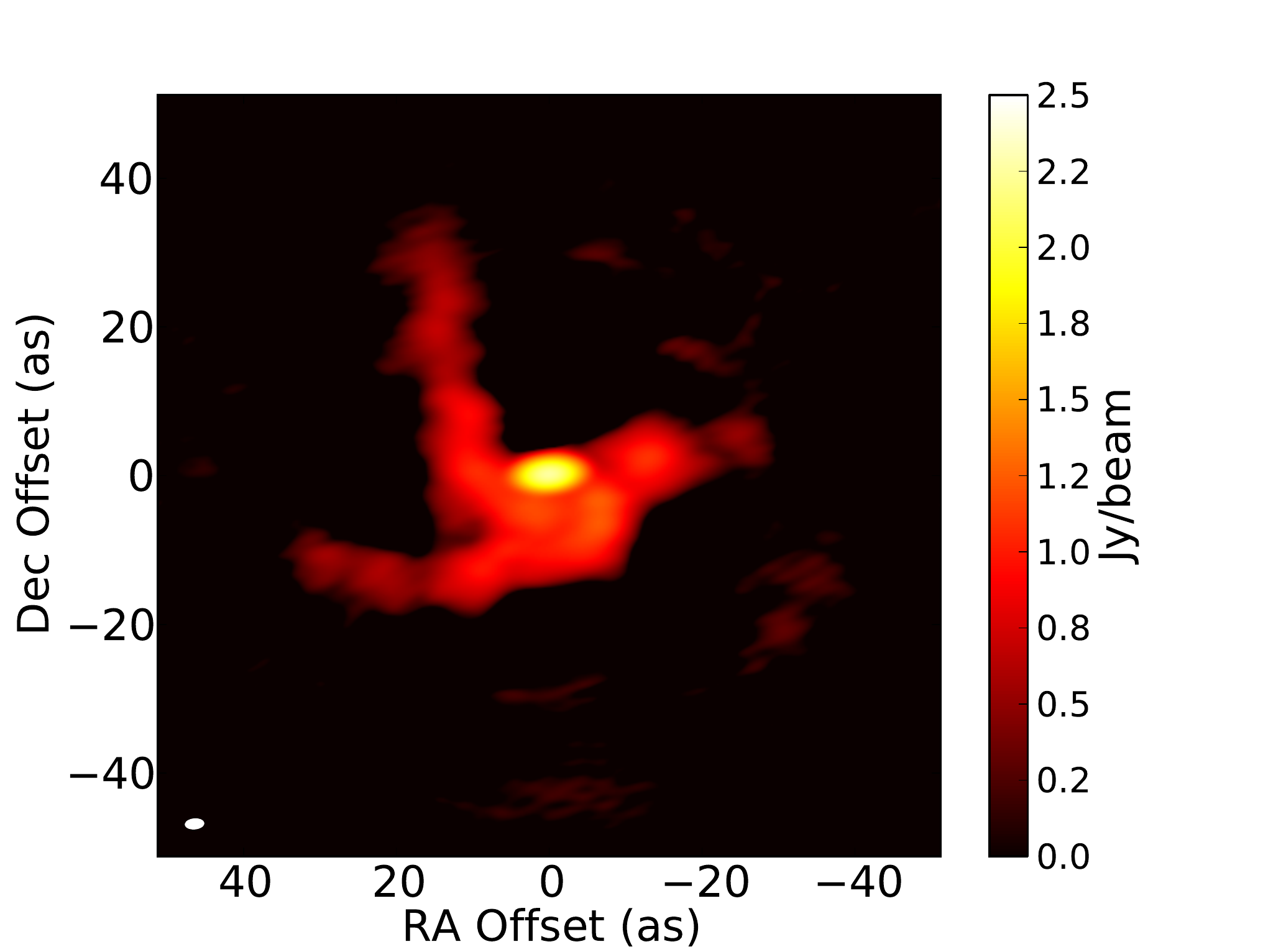}
    \caption{Image of the minispiral and \sgra (the compact point at the center of the image with a peak of~$\sim$2.4\,Jy) on April 6 after applying the two-components' self-calibration. The beam size is represented by the white ellipse at the bottom left of the image.}
    \label{fig:minispirals}
\end{figure}

Fig.~\ref{fig:minispirals} shows the final source structure (\sgra and minispiral) obtained after convergence of the two-components' self-calibration.
In Appendix~\ref{sec:appendix_calplot} the plot of the gains in phase/amplitude for each day can be found.

\section{Strategy for finding transients}
\label{sec:stgy}
To determine the presence of transients in the observations, we need to establish a criterion or a threshold in which we minimize false detection. Such errors may be related to atmospheric and instrumental effects involving suboptimal weather conditions, limited sensitivity (due to short integration time), antenna shadowing, etc. Especially notorious are the beginning and end of the Phasing System subscans, when the ALMA phase corrections have not converged yet \citep[see][]{APPRef}. We have flagged such points following~\cite{CiriacoRefb}. Final parts of the experiments are also pathological, when the antenna elevations are low and, as a consequence, the limitations due to the atmospheric opacity and turbulence are more severe. For example, the first plot on the first row of Fig.~\ref{fig:scatter_time_dynmr_11_synth} shows the dispersion in the edges of the scan during the track.

Some expected time duration for radio transient emissions are given in \cite{Lazio08}, \cite{Chiti16} and \cite{Eatough21}. While~\cite{Lazio08} presents those times for different kinds of radiosources, as ultra-high energy particles, masers and flare stars, \cite{Eatough21} refers to the pulsars and magnetar PSR J1745-2900 near the GC.  These estimates, as well as the statistics presented in~\cite{Chiti16}, are similar to 6$-$7 times the instantaneous root-mean-square (rms) of our ALMA observations, ranged between $2.138-7.210$\,mJy/beam in April 6, $0.287-16.602$\,mJy/beam in April 7 and $2.522-15.661$\,mJy/beam in April 11 (visibilities with an extremely high and nonrealistic rms have been flagged). We have thus set the detection threshold of transients to peak excursion of 7 times the instantaneous rms between consecutive images, which also gives a very low probability of false detection from thermal noise. Supposing that noise follows a normal distribution, we have computed the number of beams that form the image Fig.~\ref{fig:minispirals} (which is $\approx10^4$), and we have found that the probability of getting at least one false positive transient detection in the field of view with a confidence of 7$\sigma$ is $\approx 10^{4}\times 10^{-12}= 10^{-8}$.
\\

In addition to the detection threshold, we have assumed that the transient duration may be longer than the correlator's integration time, which implies that any source detection should be registered at several consecutive observing times. In particular, the transient search has been performed considering two different time-search intervals, one defined by the observation cadence time (4\,s) and the other one by three times the cadence time (12\,s). Excursions of at least three standard deviation in the maxima between two consecutive time-search intervals are considered as a potential transient. Therefore, on the one hand the minimum duration of a transient would be three times the integration time. On the other hand, since the minispiral model is considering all days, the maximum duration of a transient event is bounded by the track duration and the rms criterion. That means, if a transient stay for hours, we would see a compact component appearing in the residual images and disappearing after its intrinsic duration. The potential detections are then manually inspected in the residual images to ensure physical feasibility, (i.e. that they are not due to artifacts in the image plane). The times with not realistic dynamic range values are removed from the analysis. Such time-search intervals are defined following running median window at each epoch. The choice of the median is not arbitrary: while the mean would bias the window containing a very large excursion, 
 the median is better suited for skewed distributions. 
In addition, one important assumption we have made is that we are working on a Gaussian distribution of the noise. 
This is not true, as as we are dynamic range limited in snapshots where potential excursions are seen. 

Finally, to find transients in the FOV, we need to solve for the minispiral model in the image domain $\fmini$ in each different epoch separately. We use the CLEAN algorithm \citep[e.g.,][]{Hogbom1974} implemented in the Common Astronomy Software Applications (CASA) framework v5.7\footnote{https://casa.nrao.edu/}
as the \texttt{tclean} task, iteratively minimizing the difference of the residuals and the model by recalibrating the data with $G_t$, and updating $\mathcal{V}^{\mathrm{obs}}_t$. An important point to highlight is that, with this strategy, the primary-beam response is not corrected. In the case a transient would be found, the primary-beam correction should be applied. The full width half-maximum (FWHM) for ALMA at 230\,GHz is 0.5.

\section{Simulated data}
\label{sec:synthetic_data}
We have simulated ALMA observations of \sgra corresponding to the array configuration, and exact observing times, of the true visibilities of Track B (April 6) in the 2017 EHT campaign by using the CASA task \texttt{simalma} with its default parameters for 230\,GHz. Our frequency channels have a width of 2GHz, which is less than 1\

\begin{figure}
    \centering
    \includegraphics[scale=0.4]{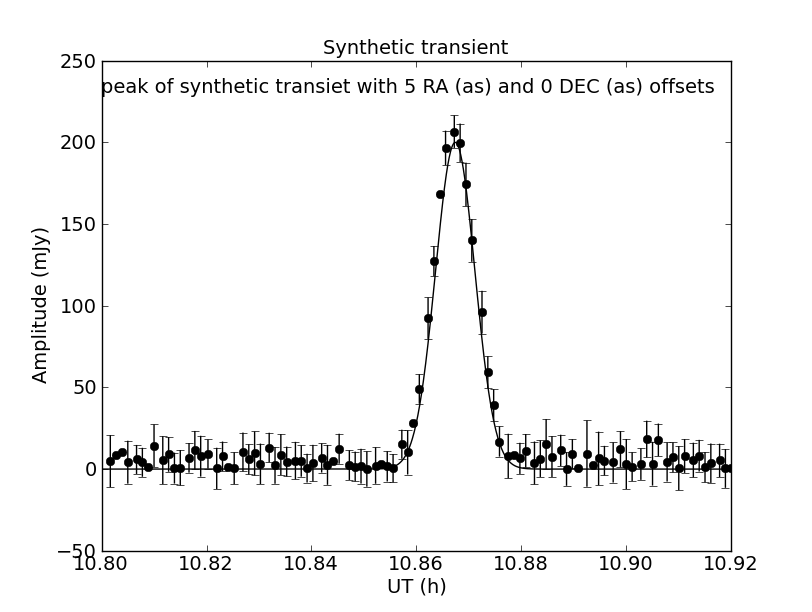}
    \caption{Amplitude as a function of time for the simulated transient. Continuous line is the simulated transient flux density. Dots are the recovered amplitude of the transient, after applying our self-calibration algorithm. Error bars correspond to the instantaneous snapshot rms}
    \label{fig:trans_ampl}
\end{figure}

In these synthetic observations, we have inserted a flat spectrum compact central source (with flux densities at each integration time, or light curves, equal to those of real \sgra) and an extended source given by our minispiral model. In addition to this, we have inserted a short transient, with amplitude evolution Gaussian-like, with a duration of $\sim42$ seconds and centered on 10:52:01 UT and located at 5 arcsec to the east of \sgra and 1 arcsec to the north. The peak flux density of the transient is $\sim207$\,mJy~(see Fig.~\ref{fig:trans_ampl}).

Thermal noise and antenna gains have been introduced based on the noise and gains found in the real data. We have calibrated the synthetic observations using the two-components' self-calibration explained in Sect~\ref{sec:stgy}. Once convergence of the self-calibration was achieved, we computed the residual visibilities (i.e., by subtracting the model of the minispiral and \sgra) and generated the corresponding residual images for each integration time. Then, the procedure described in Sect. 4 was applied to the full set of residual images in search for the transient.

\begin{figure*}
\centering
\textbf{Synthetic data}\par\medskip
\includegraphics[scale=0.4]{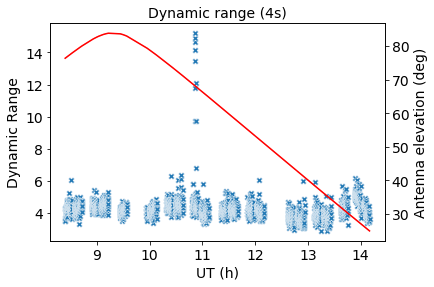}
\hspace{0.1cm}
\includegraphics[scale=0.4]{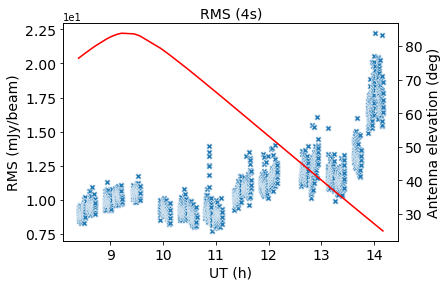}
\hspace{-0.2cm}
\\
\includegraphics[scale=0.4]{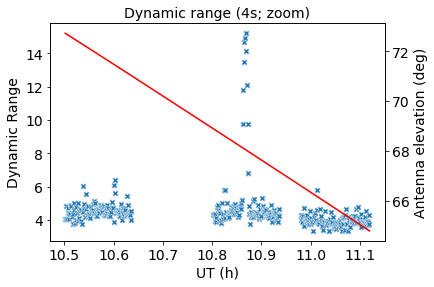}
\hspace{0.1cm}
\includegraphics[scale=0.4]{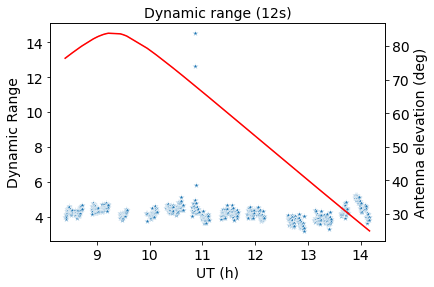}
\hspace{-0.2cm}

\caption{Snapshot image statistics for the synthetic dataset. Top left panel: time dependence of the dynamic range (with a 4\,s cadence). Bottom left panel: zoom to the UT range with the mock transient. Top right panel: RMS with a cadence of 4\,s. Bottom right panel: dynamic range using a cadence of 12\,s. In all panels, antenna elevations are shown in red.}
\label{fig:scatter_time_dynmr_11_synth}
\end{figure*}
In Fig.~\ref{fig:scatter_time_dynmr_11_synth} (top left) we show, the time evolution of the dynamic range or signal-to-noise ratio (S/N) of the residual images in the synthetic observations with a cadence of 4\,s. 
In this figure, a clear peak around 11 UT can be seen. In the top-right panel of Fig~\ref{fig:scatter_time_dynmr_11_synth}, we show the rms of the residual images.
This plot shows that the visibilities around 14 UT have higher rms than expected. The antenna elevations are shown in red as part of each of the panels in Fig~\ref{fig:scatter_time_dynmr_11_synth}. Low antenna elevation raises the rms of the signal, and so any excursions when the elevation is low are likely to be false positives. The high rms around 14 UT is therefore likely to be an effect of the low antenna elevation at the end of the experiment, not a hint of a transient. In the bottom-left panel, we show the S/N for a zoom into the time range around 11 UT, where the synthetic transient has been introduced. In order to check whether this peak could be introduced by a spurious instrumental or atmospheric event, we show, in the bottom-right panel of Fig.~\ref{fig:scatter_time_dynmr_11_synth}, the time dependence of the residual dynamic range when a time cadence of 12\,s is used to build the set of residual images.  Even in this plot, we can still see the outstanding peak in S/N around 11 UT. Hence, there is a strong evidence of a transient. As a final check, we manually inspect the residuals of the CLEAN images in that specific time. 
\label{sec:appendix_plots}
\begin{figure*}
\centering
\includegraphics[scale=0.25]{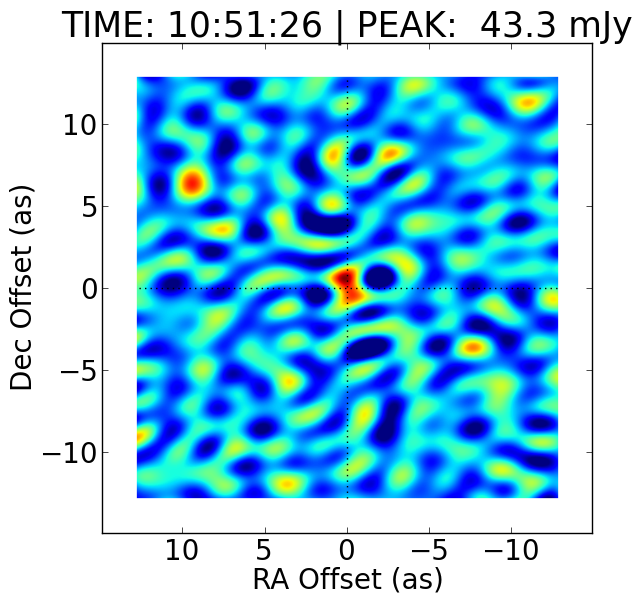}
\hspace{-0.2cm}
\includegraphics[scale=0.25]{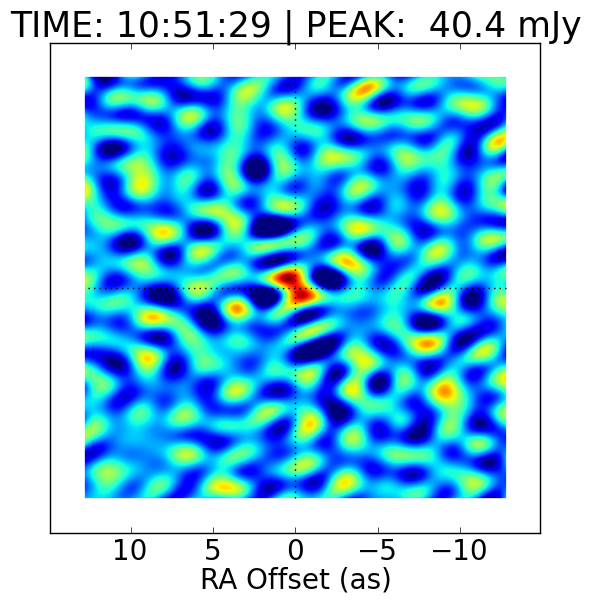}
\hspace{-0.2cm}
\includegraphics[scale=0.25]{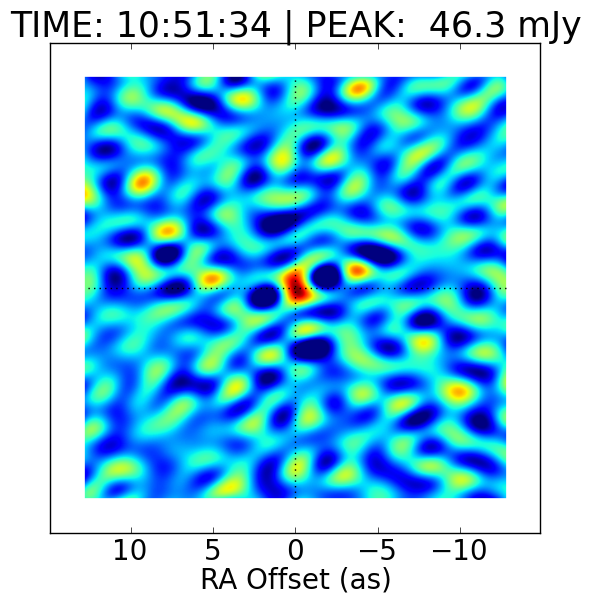}
\hspace{-0.2cm}
\includegraphics[scale=0.25]{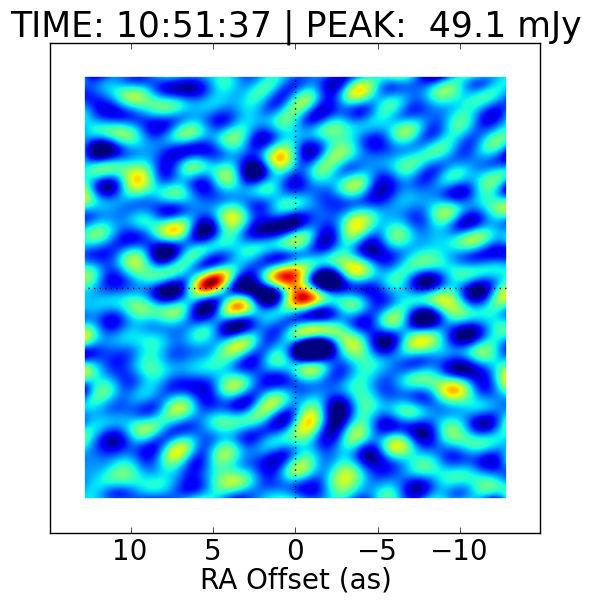}
\\
\includegraphics[scale=0.25]{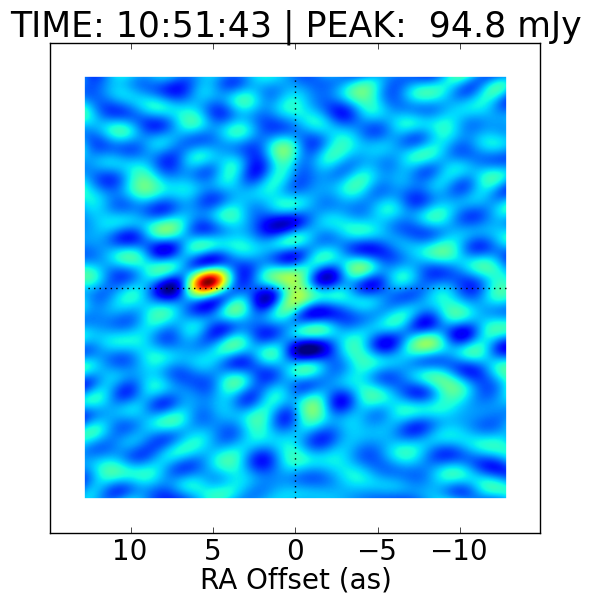}
\hspace{-0.2cm}
\includegraphics[scale=0.25]{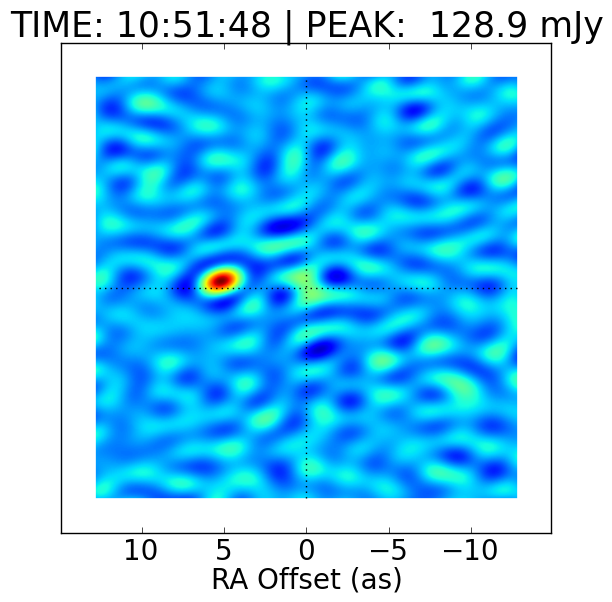}
\hspace{-0.2cm}
\includegraphics[scale=0.25]{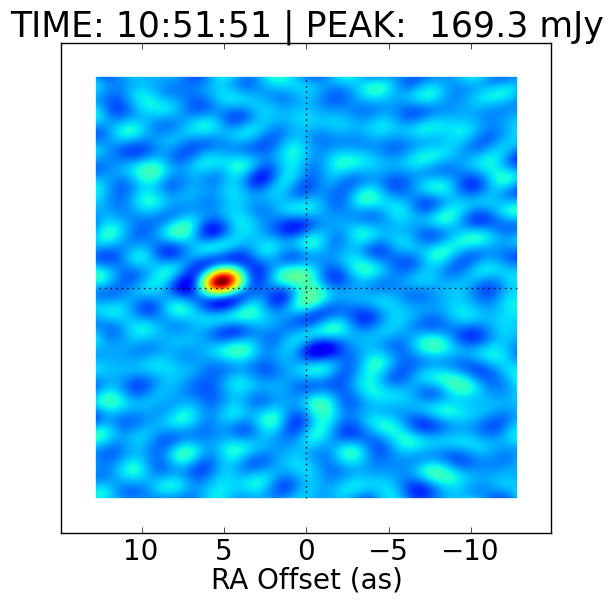}
\hspace{-0.2cm}
\includegraphics[scale=0.25]{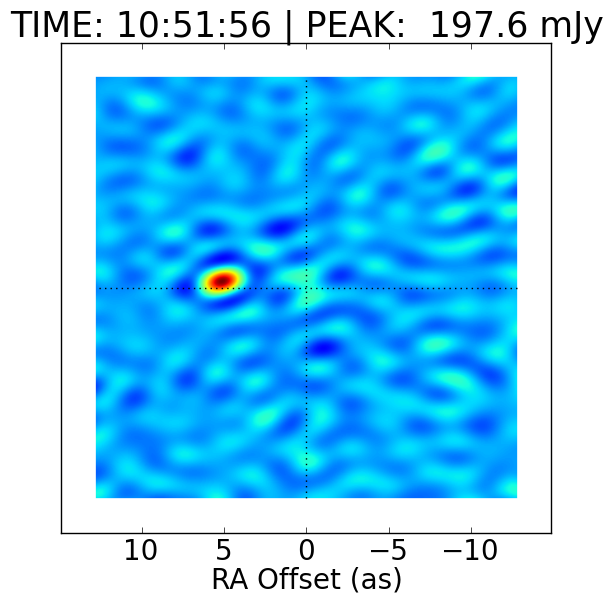}
\\
\includegraphics[scale=0.25]{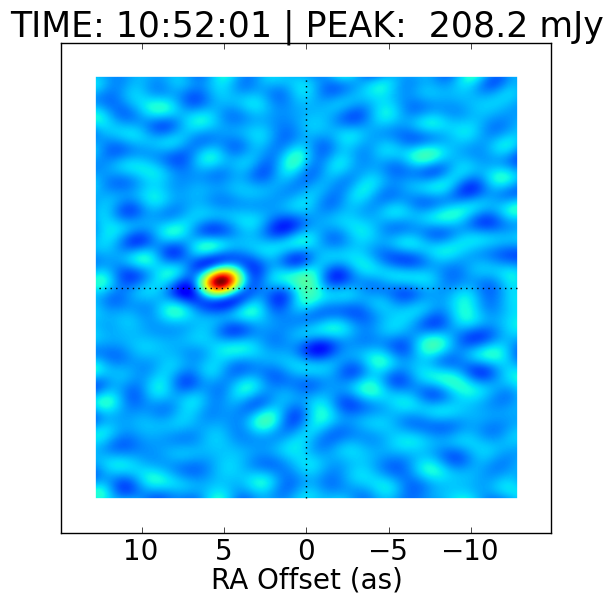}
\hspace{-0.2cm}
\includegraphics[scale=0.25]{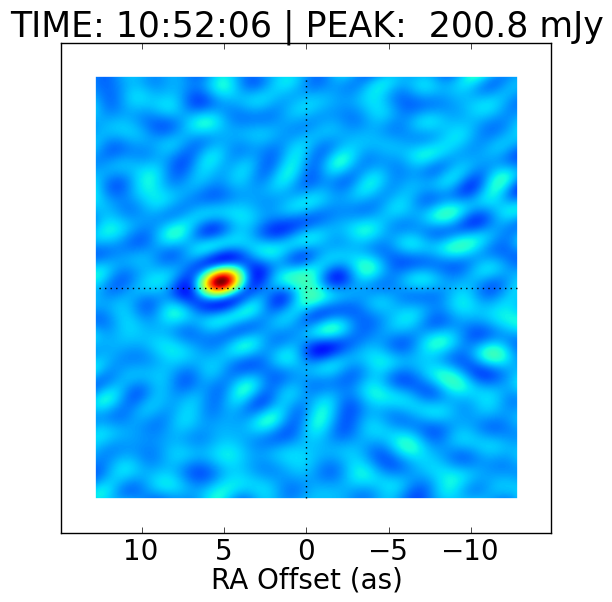}
\hspace{-0.2cm}
\includegraphics[scale=0.25]{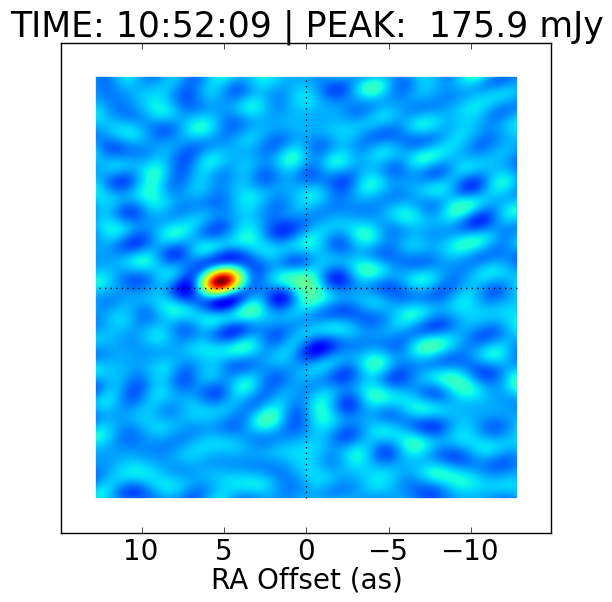}
\hspace{-0.2cm}
\includegraphics[scale=0.25]{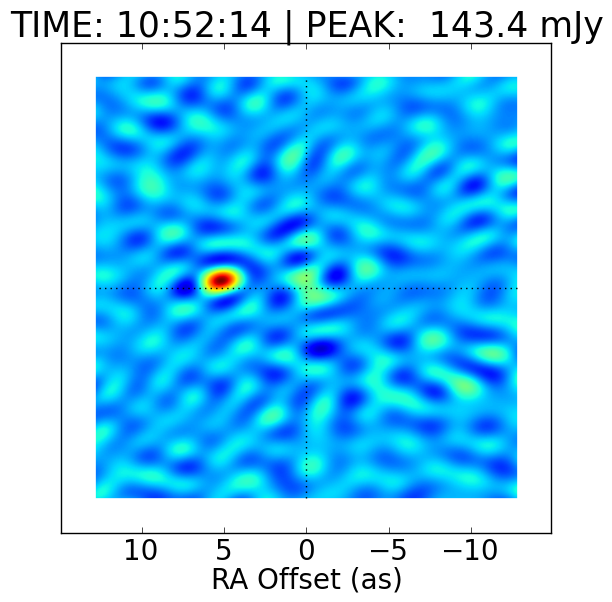}
\\
\includegraphics[scale=0.25]{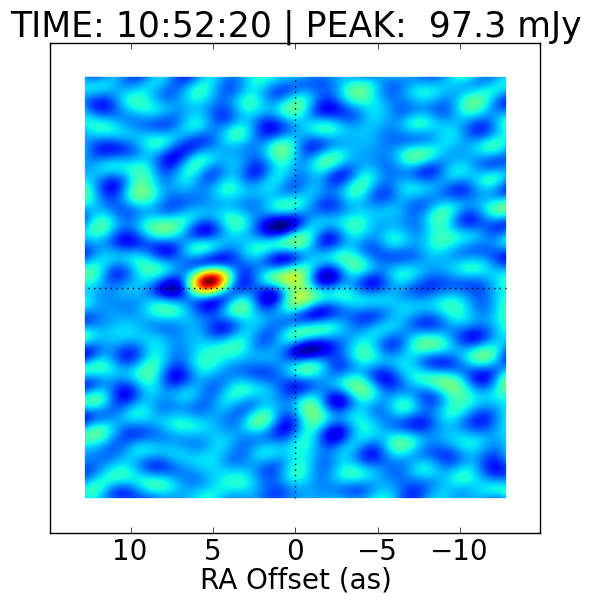}
\hspace{-0.2cm}
\includegraphics[scale=0.25]{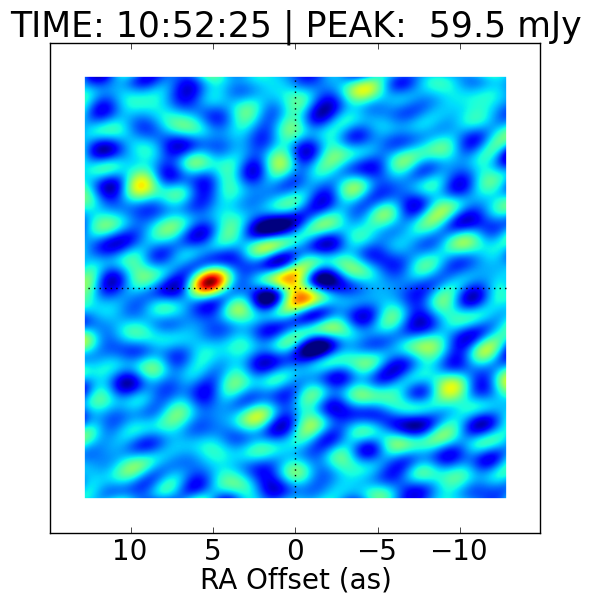}
\hspace{-0.2cm}
\includegraphics[scale=0.25]{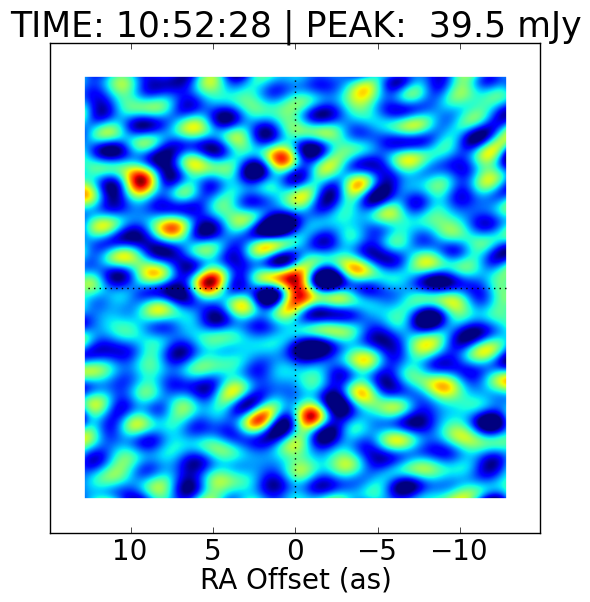}
\hspace{-0.2cm}
\includegraphics[scale=0.25]{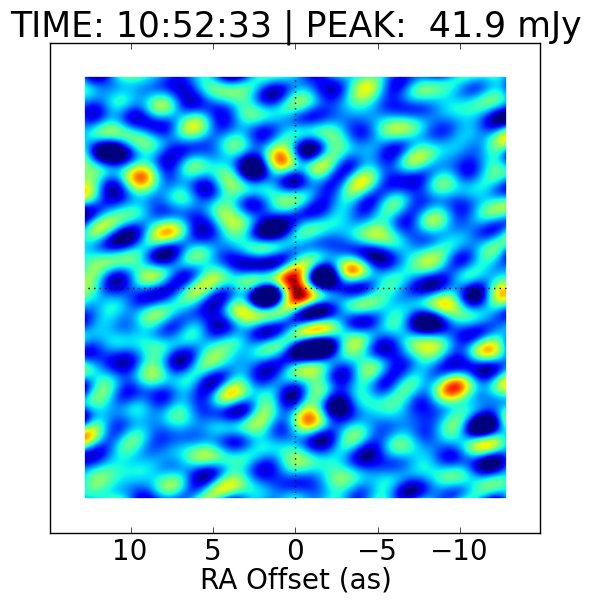}
\caption{Image residuals of synthetic data. We can see how a transient is clearly appearing, increasing remarkably the peak from~$\sim$ 43\,mJy up to~$\sim$ 200\,mJy. 
Furthermore, this transient stays more than 12\,s.}
\label{fig:clean_residuals_synth}
\end{figure*}
In Fig.~\ref{fig:clean_residuals_synth} we show residual images from several snapshots around the time of the synthetic transient. From time 10:51:27 UT (h) to 10:52:33 UT (h) we can see how a transient point source is appearing $\sim5$ arcsec to the East of \sgra. The brightness peak is notably increasing to reach a maximum of $\sim$ 200\,mJy, while the duration of the transient is approximately to 1 min. For further details, see Appendix~\ref{sec:appendix_synth}.

\section{Real data}
\label{sec:realdata}
As explained in Sect.~\ref{sec:observations}, there were three days of observations, April 6 (track B), April 7 (track C), and April 11 (or track E). For each day the data have been calibrated using the two-component' self-calibration technique described in Sect.~\ref{sec:stgy}.\\

Another interesting point to note is that the rms of the real data is lower than the rms of the synthetic one. This could be due to the fact that real observing conditions are less sever than the one considered by the CASA task \texttt{simalma} used to simulate the ALMA visibilities.

\begin{figure*}
\centering
\includegraphics[scale=0.3]{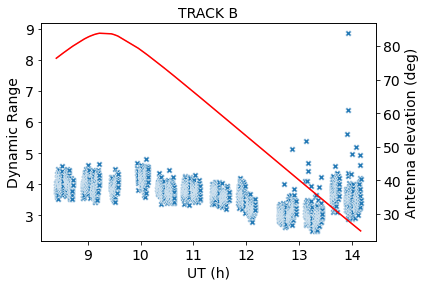}
\hspace{-0.1cm}
\includegraphics[scale=0.3]{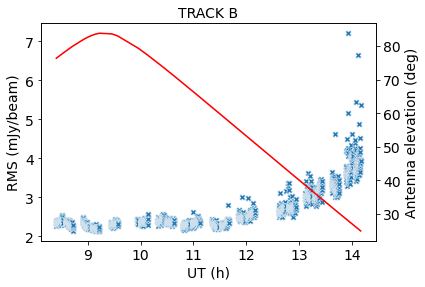}
\hspace{-0.3cm}
\includegraphics[scale=0.3]{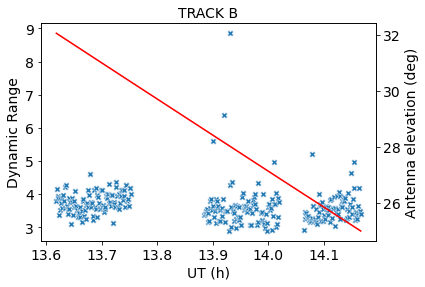}
\hspace{-0.2cm}
\includegraphics[scale=0.3]{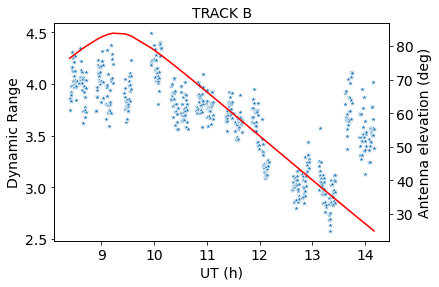}
\\
\includegraphics[scale=0.3]{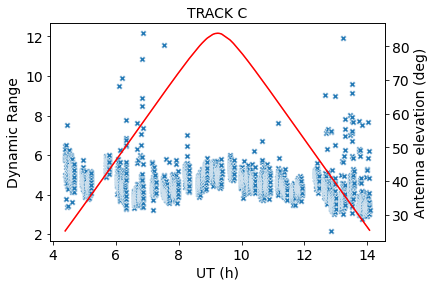}
\hspace{-0.2cm}
\includegraphics[scale=0.3]{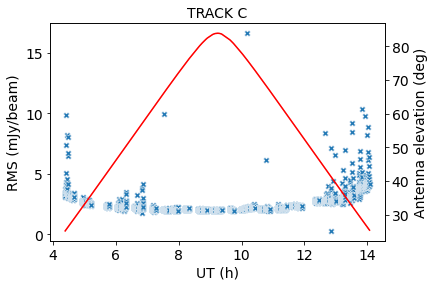}
\hspace{-0.2cm}
\includegraphics[scale=0.3]{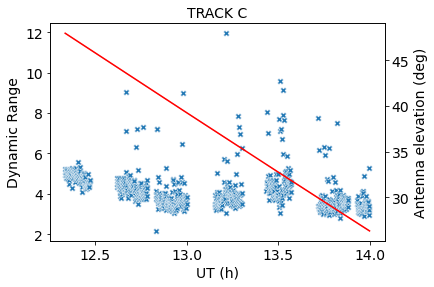}
\hspace{-0.2cm}
\includegraphics[scale=0.3]{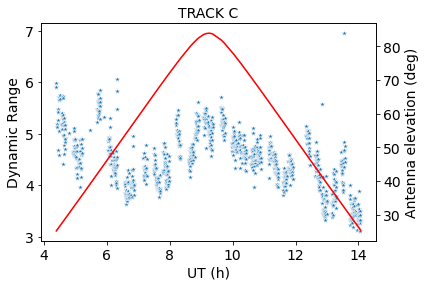}
\\
\includegraphics[scale=0.3]{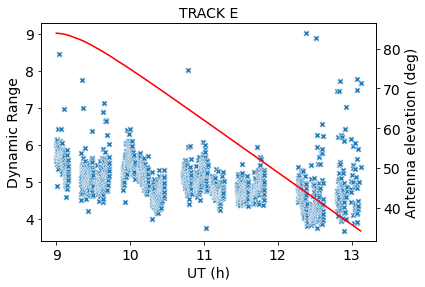}
\hspace{-0.2cm}
\includegraphics[scale=0.3]{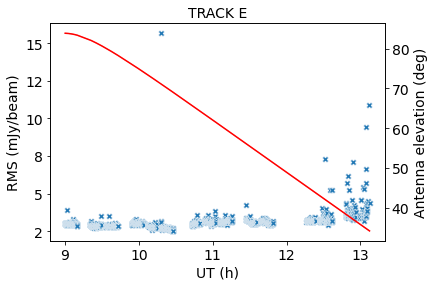}
\hspace{-0.2cm}
\includegraphics[scale=0.3]{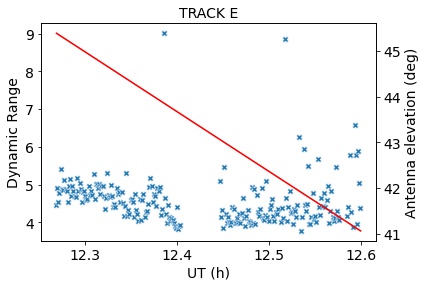}
\hspace{-0.2cm}
\includegraphics[scale=0.3]{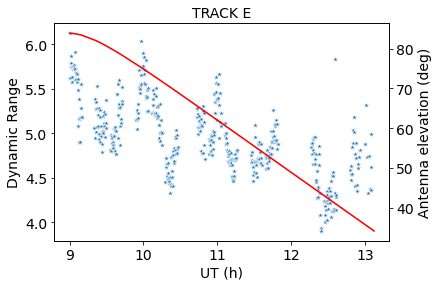}

\caption{Snapshot residual image statistics after the two-components' self-calibration. Each row refers to one epoch (see Sect.~\ref{sec:observations} for details). Columns from left to right: dynamic range of the experiments (4\,s cadence); rms of the experiments (4\,s cadence); zoom to the area where more promising excursions have been detected using 4\,s cadence; dynamic range (12\,s cadence). In all panels, antenna elevations are shown in red.}

\label{fig:scatter_time_dynmr_11}
\end{figure*}

Fig.~\ref{fig:scatter_time_dynmr_11} shows the dynamic range and the rms of each of the residual snapshot images along the experiment, together with the antenna elevations. We can observe some excursions that are close to the threshold limit, counting as transient candidates. It is important to note that data corresponding to source elevation below 25$^\circ$ is exhibiting significant quality loss. Each column in Fig.~\ref{fig:scatter_time_dynmr_11} represents 

the dynamic range and rms during the observation for 4\,s and 12\,s and a zoom in a promising transient detection area, while each row corresponds to a different track (from top to bottom: track B, track C and track E). All the transient candidates (seen as outliers in the S/N time evolution) are excluded as true transients, either because they are only seen in one correlator integration time (4\,s), or because they fail the threshold criterion (see Sect.~\ref{sec:stgy}). Nevertheless, we note that the residual snapshot images in April 7 are more variable in dynamic range than the rest of the epochs. Even, some of the excursions survive the 12\,s integration time. Such track is especially bad, in particular at the beginning and at the end of the experiment. In Appendix~\ref{sec:appendix_plots}, we show the residual images corresponding to the observing times of April 7 when potential transients would be detected, based on the selection criteria. The systematic spatial distribution of image residuals in these frames suggests a high dynamic-range limitation, related to a suboptimal calibration possibly related to especially bad weather. \\

Finally, we show in Tables~\ref{tab:trackb}, \ref{tab:trackc} and \ref{tab:tracke} (top rows) the dynamic range, rms, UT epoch, and antenna elevation of the residual images with highest dynamic ranges. 
The bottom table rows show the mean of the dynamic range and the mean and the std of the rms of the residual images. These tables are based on Fig.~\ref{fig:scatter_time_dynmr_11} and by manual inspecting the residuals. Then choosing those frames with higher S/N. 
\\
In conclusion, in all cases, the peaks or excursions of the dynamic range are below the threshold limits. Thus we can conclude that we have not found any reliable detection of transients in the ALMA observation during the 2017 EHT Campaign. In any case, in Appendix~\ref{sec:appendix_plots}, detailed sequences of snapshots of transient candidates are shown. The aim of these plots is to verify whether the higher dynamic range in a particular UT is related to a point source in the image plane, or if the image is contaminated by systematic artifacts, which would be related to instrumental and/or atmospheric residual effects.

\begin{table*}
\caption{Peak dynamic range of the residual images during the track B.}
\centering

\begin{tabular}{rrlrrr}
\toprule
\small{      S/N} &       \small{RMS (mJy/beam)} &        \small{UT} &  \small{Ant. Elev. (deg.)} &      \small{RA. (mas)} &        \small{DEC. (mas)} \\
\midrule
5.397 & 3.412 &  13:08.22 &     38.43 &      24.9 &    -8.89 \\
5.609 & 4.165 &  13:54.00 &     28.03 &      58.3 &    -24.1  \\
6.378 & 4.052 &  13:55.12 &     28.04 &      64.6 &    -28.8  \\
8.854 & 7.210 &  13:55.50 &     28.00 &      65.1 &    -33.1  \\
\bottomrule
\end{tabular}

\begin{tabular}{rrrr}
\toprule
     S/N MEAN & RMS MEAN (mJy/beam) & RMS STD (mJy/beam) \\
\midrule
 3.660 & 2.628 & 0.494 \\
\bottomrule
\end{tabular}
\\

\label{tab:trackb}

\raggedright{\textbf{Notes.} Residual images around the peak of the dynamic range have been manually inspected even if they were not satisfying the time constraint and the rms threshold.} 
\end{table*}

\begin{table*}
\caption{Peak dynamic range of the residual images during the track C.}
\centering

\begin{tabular}{rrlrrr}
\toprule
\small{      S/N} &       \small{RMS (mJy/beam)} &        \small{UT} &  \small{Ant. Elev. (deg.)} &      \small{RA. (mas)} &        \small{DEC. (mas)} \\
\midrule
 9.498 & 2.921 &   6:05.56 &     47.35 &     0.0 &    0.0 \\
 9.885 & 2.945 &   6:12.42 &     48.83 &     0.0 &    0.0 \\
10.854 & 3.837 &   6:49.25 &     57.35 &     0.0 &    0.0 \\
12.164 & 4.194 &   6:51.33 &     57.99 &     0.0 &    0.0 \\
11.581 & 9.908 &   7:32.56 &     66.37 &     0.0 &    0.0 \\
11.947 & 5.292 &  13:12.48 &     36.47 &      25.5 &     -10.1 \\
 9.575 & 8.442 &  13:30.25 &     32.69 &      26.3 &    -11.6 \\
 9.132 & 9.183 &  13:31.32 &     32.68 &      26.9 &    -18.8 \\
\bottomrule
\end{tabular}

\begin{tabular}{rrrr}
\toprule
     S/N MEAN & RMS MEAN (mJy/beam) & RMS STD (mJy/beam) \\
\midrule
 4.537 & 2.490 & 0.749 \\
\bottomrule
\end{tabular}
\\

\textbf{Notes.} For description see notes of Table~\ref{tab:trackb}.
\label{tab:trackc}
\end{table*}

\begin{table*}
\caption{Peak dynamic range of the residual images during the track E.}
\centering

\begin{tabular}{rrlrrr}
\toprule
\small{      S/N} &       \small{RMS (mJy/beam)} &        \small{UT} &  \small{Ant. Elev. (deg.)} &     \small{RA. (mas)} &        \small{DEC. (mas)} \\
\midrule
8.444 &  3.897 &   9:01.54 &     83.84 &     -3.7 &    -1.8 \\
7.742 &  3.206 &   9:20.41 &     78.86 &     -3.4 &    -1.8 \\
8.007 &  3.572 &  10:47.02 &     64.89 &      0.0 &       0.0 \\
9.008 &  3.432 &  12:23.08 &     43.86 &      0.0 &    0.0 \\
8.867 &  7.299 &  12:31.02 &     42.33 &     0.75 &    -0.25 \\
7.728 &  5.232 &  12:50.31 &     37.99 &      0.5 &      -0.5 \\
7.787 &  9.429 &  13:04.15 &     34.95 &     2.2 &    -1.41 \\
7.659 & 10.870 &  13:07.17 &     34.26 &     1.78 &    -1.7 \\
\bottomrule
\end{tabular}
\begin{tabular}{rrrr}
\toprule
     S/N MEAN & RMS MEAN (mJy/beam) & RMS STD (mJy/beam) \\
\midrule
    5.005 & 3.096 & 0.521 \\
\bottomrule
\end{tabular}
\\
\textbf{Notes.} For description see notes of Table~\ref{tab:trackb}.

\label{tab:tracke}
\end{table*}

\section{Summary and conclusions}
\label{sec:summary}

The GC is a turbulent region with a relatively high density of gas, dust and stars. As a consequence, different astrophysical phenomena, related to transit emission, may occur in this region.

The superb sensitivity of ALMA and the high time cadence of the ALMA-VLBI observations of the GC performed during the 2017 EHT campaign, allowed us to perform a search down to second/minute scale radio transients in the GC field at arc-second resolution. Even with integration times as short as 4\,s, the ALMA images still show clear signatures of dynamic-range limitation, which indicates that the thermal noise sensitivity level is not reached.

The supermassive black hole \sgra strongly dominates the field of view of ALMA and, thus, the image of the minispirals can contain large sidelobe effects (related to the intrinsic variability of SgrA*) that may lead to spurious transient detections. For this reason, a careful calibration and subtraction of the \sgra signal must be performed, prior to the transient search. In this work we present a nonstandard calibration procedure for a source that can be modeled using two different components: a compact and variable source and a second extended one, that is characterized by a constant brightness distribution. This technique only requires to know the source structure and to detect the whole flux. The brightness intensity is not relevant, as long as previous hypothesis are satisfied~\citep[see e.g.][]{Radcliffe16}.
This strategy allows us to estimate antenna gains accurately 
and get more precise information on the light curves of the compact component in a way similar to what is presented in~\cite{Wielgus21}. Using this calibration procedure, we have performed a search for transients in the GC from ALMA observations obtained during during the 2017 EHT campaign (April 7, April 9 and April 11), by tracking events occurring for a duration of at least 12s (three times the ALMA integration time). Such events also had to satisfy the criterion that the peak of the dynamic range between consecutive images is about 7 times the rms of the observation. These criteria have been established based on both on statistical arguments (i.e., acceptably low probability of false detection from thermal noise) and on previous transient searches reported at other wavelengths.

Using the criteria defined in section~\ref{sec:stgy}, we have not observed any transient event with sufficient statistical significance. However, the new two-source self-calibration applied to future ALMA observations will allow us to perform accurate, both in time and astrometic, variability analyses of the GC field with unprecedented sensitivity increasing the probability of transients detection.

\begin{acknowledgement}
This work has been partially supported by the MICINN Research Project PID2019-108995GB-C22. IMV and AM also thank the Generalitat Valenciana for funding, in the frame of the GenT Project CIDEGENT/2018/021. 
The authors thank Michael Janssen and Geoff Bower for their valuable comments to improve this work.
We thank the referee for their very careful and deep review of the paper, and for the comments, corrections and suggestions.
\end{acknowledgement}

\clearpage
\appendix
\input{Appendix_math}
\input{Appendix_calibration_plots}
\input{Appendix_synth}
\input{Appendix_plots}

\end{document}

%% file: Appendix_math.tex
\section{Mathematical details of the two-components' self-calibration}
\label{sec:appendix}

In this Appendix, we present the mathematical details of the two-components' self-calibration strategy.
Eq.~\ref{eq:model} can be seen as four-dimensional fitting whose variables are $\fmini, \fsgra, \deltara, \deltadec$.
Then, the fitting minimization problem can be solved applying a simple Newton method, whose recursive formulation reads

\begin{equation}
\label{eq:app_newton}
    \begin{pmatrix}
    \fmini\\
    \fsgra\\
    \deltara\\
    \deltadec\\
  \end{pmatrix} =
  \mathcal{H}^{-1}
    \begin{pmatrix}
    R_{0}\\
    R_{1}\\
    R_{2}\\
    R_{3}\\
  \end{pmatrix},
\end{equation} 

where $\mathcal{H}$ is the Hessian of the $\chi^2$ with respect to each parameter and the components $R_k, k=1,\ldots,4$ are its derivatives. Without loss of generality, and for an easier readability, we can do all the analysis fixing a given baseline $j$ and omitting this subindex.\\
\\

Let $H_{\text{mn}}, m,n=1,\ldots,4$ be the entries of the $\mathcal{H}$ matrix. Eq.~\eqref{eq:app_newton} is

\begin{equation}\label{eq:extended_app_newton}
    \begin{pmatrix}
    \fmini\\
    \fsgra\\
    \deltara\\
    \deltadec\\
  \end{pmatrix} =
  \begin{pmatrix}
    H_{00} & H_{01} & H_{02} & H_{03}\\
    H_{10} & H_{11} & H_{12} & H_{03}\\
    H_{20} & H_{21} & H_{22} & H_{03}\\
    H_{30} & H_{31} & H_{32} & H_{03}
  \end{pmatrix}^{-1}
    \begin{pmatrix}
    R_{0}\\
    R_{1}\\
    R_{2}\\
    R_{3}\\
  \end{pmatrix}.
\end{equation}

Several considerations are worth to note: First, the model-fitting problem is linear in $\fmini$ and $\fsgra$ while $\deltara$ and $\deltadec$ can be linearized using first order Taylor expansion. Second, $\vmod$ is complex. 
We can fit real and imaginary parts independently (for a discussion about complex-valued model fitting in interferometry see~\citep[e.g.,]{imv2014}. Finally, $\fsgra$ is a real value.\\
Let us denote by $\phi:=\dfrac{2\pi i}{\lambda}\left( u\deltara + v\deltadec\right)$ being $i$ the imaginary unit, $\phi_u:=\dfrac{2\pi i}{\lambda}u$ and $\phi_v:=\dfrac{2\pi i}{\lambda}v$. The derivatives $R_k$ can be written as

\small{
\begin{align*}
  &R_0 = 2\displaystyle{\sum_k\left(\Re\left(\vmod\right)\Re\left(\vobsnb\right)+\Im\left(\vmod\right)\Im\left(\vobsnb\right)\right) w_k}e^{\phi}; \\
  &R_1 = 2\displaystyle{\sum_k\Re\left(\vobsnb\right) w_k}e^{\phi};\\
  &R_2 = \displaystyle{\sum_k}{\Re\left(\phi_u \vmmod\right)\Re\left(\vobsnb\right) + \Im\left(\phi_u \vmmod\right)\Im\left(\vobsnb\right)};\\
  &R_3 = \displaystyle{\sum_k}{\Re\left(\phi_v \vmmod\right)\Re\left(\vobsnb\right) + \Im\left(\phi_v \vmmod\right)\Im\left(\vobsnb\right)},
\end{align*} 
}
\normalsize
being $\vobsnb$ the vector of the observed visibilities. Each entree $H_{\text{mn}}$ is of the form

\tiny{
\begin{align*}
  &H_{00} = 2\displaystyle{\sum_k\lvert \vmod\rvert^2 w_k}e^{\phi};\ H_{11} = 2\displaystyle{\sum_k w_k}e^\phi;\\
  &H_{10}=H_{01}=2\displaystyle{\sum_k}\Re\left(\vmod\right)w_k;\\
  &H_{02}=H_{20}=2\displaystyle{\sum_k\left(\Re\left(\vmod\right)\Re\left(\vmmod\phi_u\right)+\Im\left(\vmod\right)\Im\left(\vmmod\phi_u\right)\right)w_k}\\
  &H_{03}=H_{30}=2\displaystyle{\sum_k\left(\Re\left(\vmod\right)\Re\left(\vmmod\phi_v\right)+\Im\left(\vmod\right)\Im\left(\vmmod\phi_v\right)\right)w_k}\\
  &H_{12}=H_{21}=2\displaystyle{\sum_k \vmmod\phi_u w_k}e^\phi;\\
  &H_{13}=H_{31}=2\displaystyle{\sum_k \vmmod\phi_v w_k}e^\phi;\\
  &H_{22}=2\displaystyle{\sum_k\lvert \vmmod\phi_u\rvert^2 w_k}e^{\phi};\\
  &H_{23}=H_{32}=2\displaystyle{\sum_k \vmmod\left(\Re\left(\phi_u\right)\Re\left(\phi_v\right)+\Im\left(\phi_u\right)\Im\left(\phi_v\right)\right) w_k}e^\phi;\\
  &H_{33}= 2\displaystyle{\sum_k\lvert \vmmod\phi_v\rvert^2 w_k}e^{\phi}.
\end{align*} 
}
\normalsize

In this way, we have computed all terms of for solving the linear system~\ref{eq:extended_app_newton}.

%% file: Appendix_calibration_plots.tex
\section{Calibration plots}
\label{sec:appendix_calplot}

In this Appendix, we show the distributions of antenna gains resulting from the two-components' self-calibration. In the case of the real data, these gains are incremental with respect to those from the ALMA-VLBI QA2 calibration~\citep[see][]{CiriacoRefb}, which implies that they should be close to unity (i.e., with a phase close to zero and an amplitude close to 1).

\begin{figure*}
\centering
\includegraphics[scale=0.6]{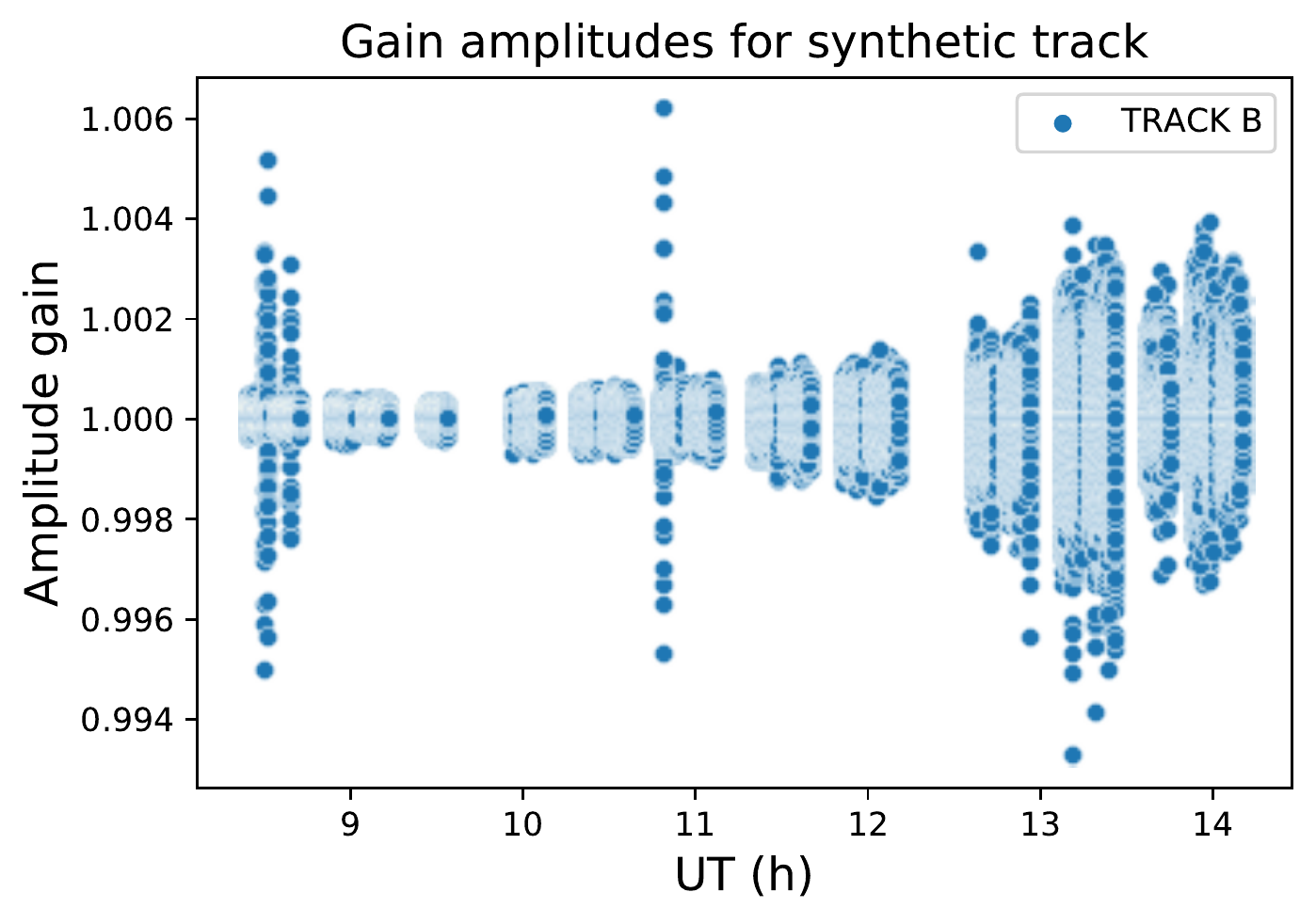}
\hspace{-0.0cm}
\includegraphics[scale=0.6]{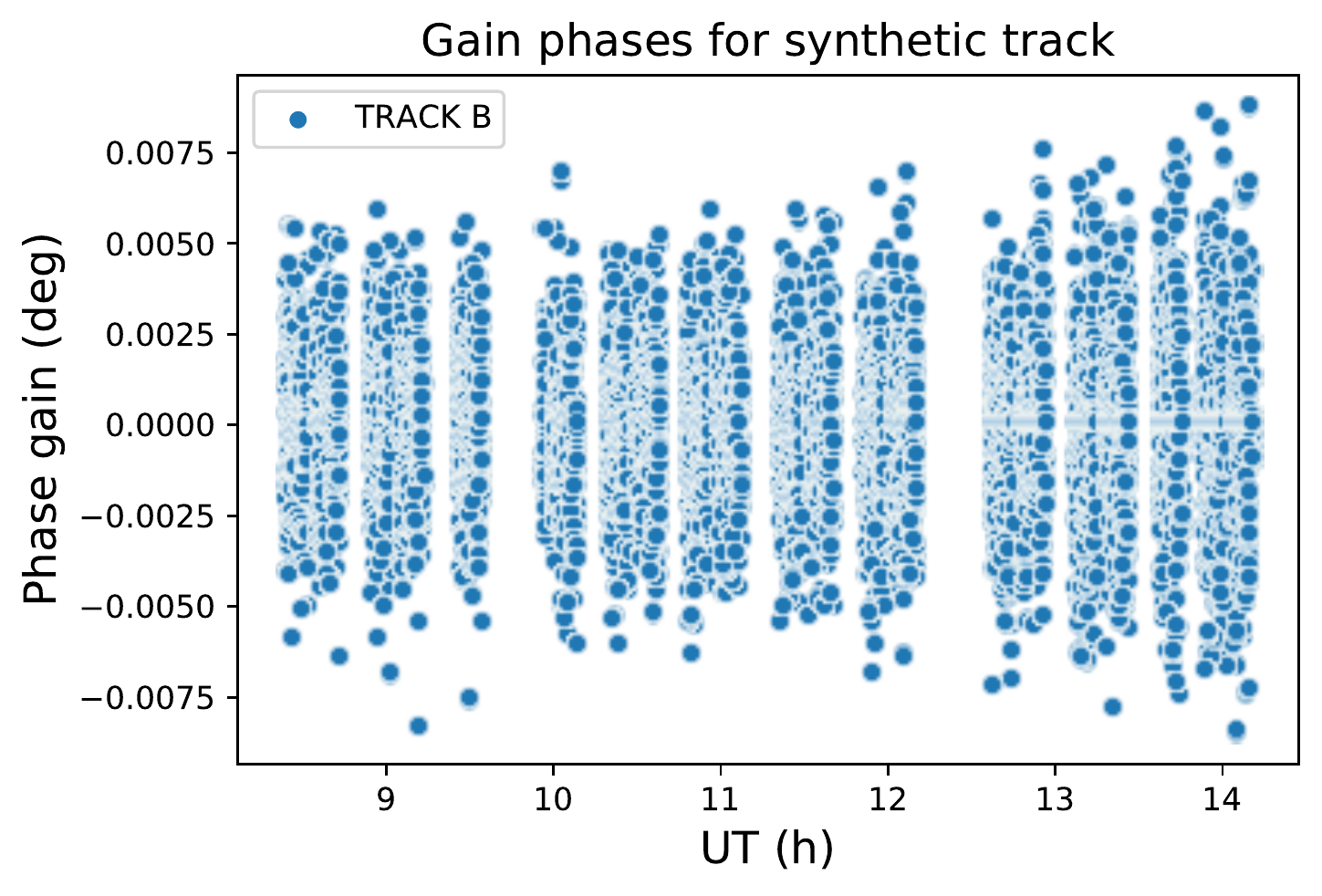}
\caption{Left panel: gains in amplitude vs UT of the synthetic track. Right panel: same as left, but phase gains.}
\label{fig:gains_fake}
\end{figure*}

Figure~\ref{fig:gains_fake} shows the gains in phase and amplitude vs time for the synthetic data. Note that the amplitude gains are all around 1.
\begin{figure*}
\centering

\includegraphics[scale=0.6]{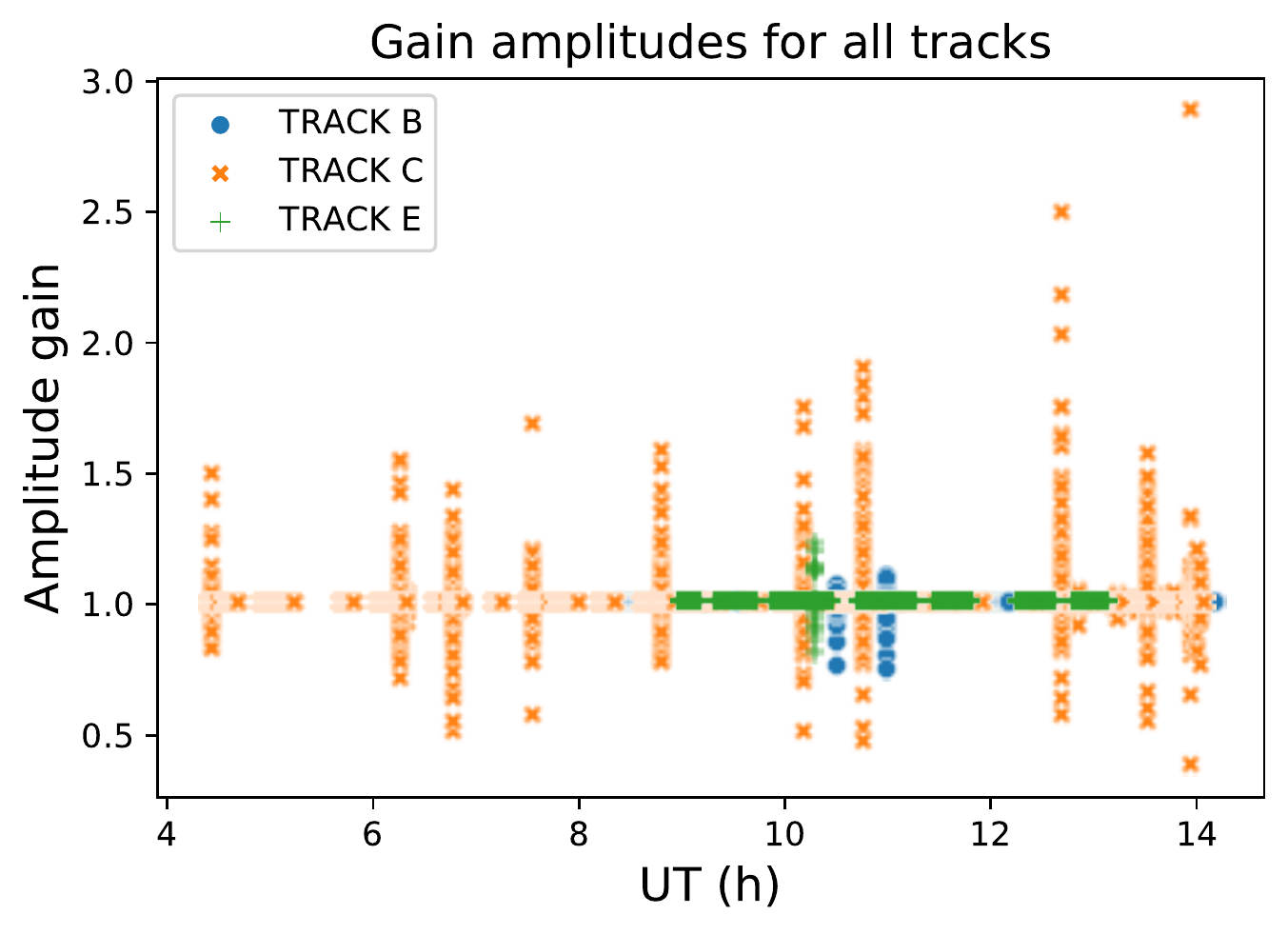}
\includegraphics[scale=0.6]{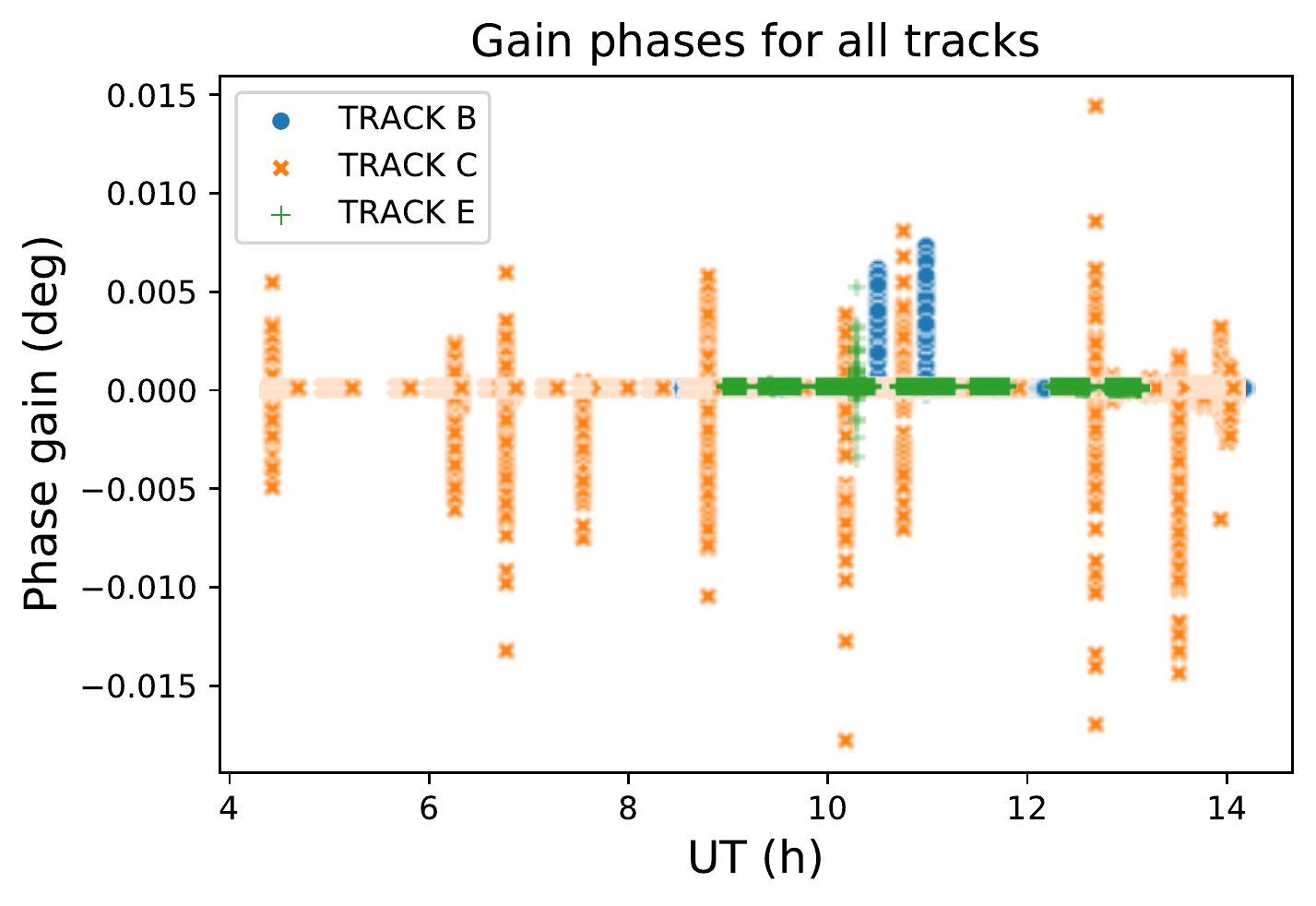}
\caption{Left panel shows the amplitude gains for different UT. Right panel, the phase gains. Each track has different color and different marker.}
\label{fig:gains_real}
\end{figure*}
Figure~\ref{fig:gains_real} is the equivalent of the former but for the real data. Different tracks are represented by different markers and colors. 
In all cases, we see that the incremental gains are actually close to unity, as expected.

\begin{figure*}
    \centering
    \includegraphics[scale=0.4]{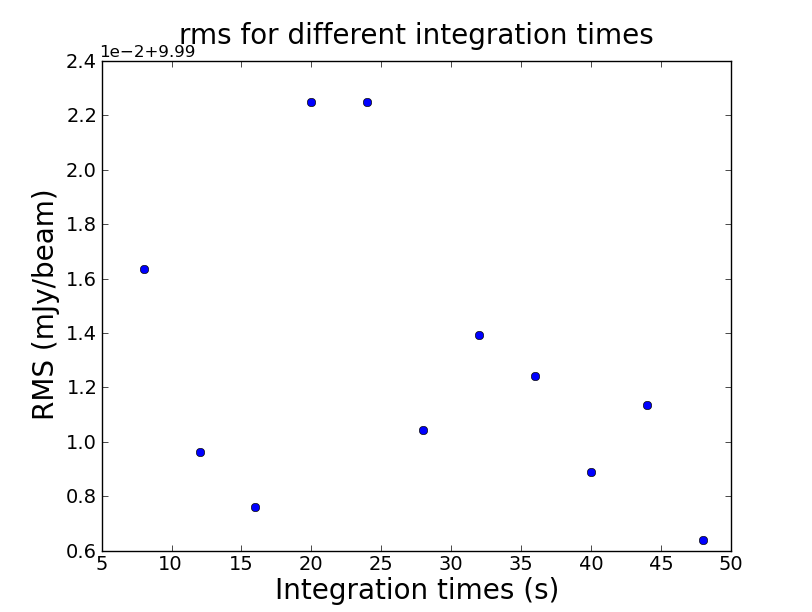}
    \includegraphics[scale=0.4]{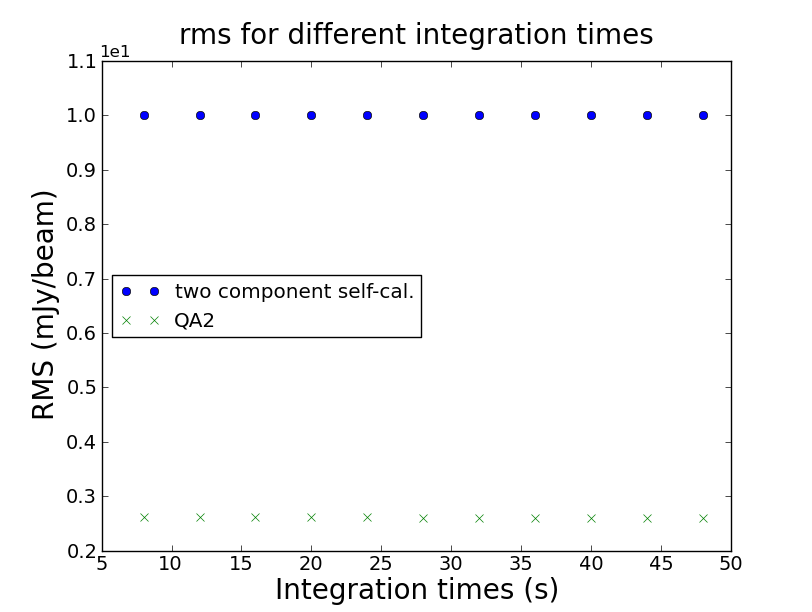}
    \caption{Rms vs time for the case of Track B (April 6). Left panel: only showing the rms for the two-components' self-calibration. Right panel: comparison between QA2 rms (marked with cross) and two-components' self-calibration (dots).}
    \label{fig:rms_vs_time}
\end{figure*}

In Figure~\ref{fig:rms_vs_time} left panel, we show the rms of the residual image (i.e., after subtracting \sgra and the minispiral) on April 6 at time 09:26:04, as a function of integration time. If the observations were limited by thermal noise, the rms should decrease as the square root of the integration time. However, the figure shows an almost flat rms, which is indicative of a limitation by dynamic range. Our analysis is, thus, limited by the contrast with respect to the total flux density of the model (i.e., \sgra and minispiral), likely due to the finite fidelity of our model (in particular, the minispiral image). In addition, Fig.~\ref{fig:rms_vs_time} right panel shows a comparison of the rms obtained by taking the QA2 data and cleaning with the same mask and \texttt{CLEAN} parameters used for the two-components' self-calibration. We can see an improvement on the rms by a factor of $\sim$5.

%% file: Appendix_synth.tex
\section{Some details on the synthetic transient}
\label{sec:appendix_synth}

In this Appendix, we present some statistics on the synthetic data described in Sect.~\ref{sec:synthetic_data}. In particular, we show the complete evolution of the transient in detail. The top of Table~\ref{tab:tracks} presents the values of the dynamical range (S/N), rms (RMS), UT and antenna elevation of the experiment. The bottom, the mean and standard deviation of the S/N and rms. To construct this table, we have identified (using Fig.~\ref{fig:scatter_time_dynmr_11_synth}) the transient period. Once we know the time interval, we can check the antenna elevation to check whether the observations may be affected by either a large air mass and/or shadowing. We can actually observe that values are bounded from $\sim$ 77 deg to $\sim$ 83 deg, which correspond to acceptable antenna elevations. Now, we can easily follow the evolution of the S/N and rms. In both cases, they start with relative low (close to the mean) and they start to increase progressively while they are advancing in time, until their maximum (15.19 , 1.32\,mJy/beam respectively) is reached. In this moment, at 10:52.05 UT, the transient reaches its peak. Then, it starts decreasing its flux density, until it becomes undetectable. We can see how, based on the bottom part of the Table~\ref{tab:tracks}, the S/N has increased during the transient with respect to the mean of the whole experiment.

\begin{table}
\caption{Peak dynamic range of the residual images in the synthetic data}
\centering

\begin{tabular}{rrlr}
\toprule
\small{      S/N} &       \small{RMS (mJy/beam)} &        \small{UT} &  \small{Ant. Elev. (deg.)} \\
\midrule
 6.076 & 10.090 &   8:30.36 &     77.64 \\
 6.303 & 10.115 &  10:24.41 &     73.86 \\
 6.037 &  9.098 &  10:32.16 &     72.17 \\
 6.117 &  8.951 &  10:36.04 &     71.47 \\
 6.367 &  9.168 &  10:36.14 &     71.44 \\
 5.813 &  8.914 &  10:49.26 &     68.61 \\
 5.812 &  9.172 &  10:49.36 &     68.58 \\
 9.742 &  9.730 &  10:51.43 &     68.12 \\
11.797 & 10.928 &  10:51.48 &     68.11 \\
13.492 & 12.550 &  10:51.51 &     68.09 \\
14.666 & 13.473 &  10:51.56 &     68.08 \\
14.921 & 13.955 &  10:52.01 &     68.03 \\
15.194 & 13.218 &  10:52.06 &     68.04 \\
14.121 & 12.455 &  10:52.09 &     68.03 \\
12.096 & 11.852 &  10:52.14 &     68.01 \\
 9.762 &  9.969 &  10:52.20 &     67.99 \\
 6.831 &  8.713 &  10:52.25 &     67.97 \\
 5.822 &  9.978 &  11:00.48 &     66.16 \\
 6.044 & 13.827 &  12:04.16 &     52.29 \\
 5.990 & 16.040 &  12:54.39 &     41.27 \\
 5.916 & 11.982 &  13:08.22 &     38.42 \\
 6.187 & 15.272 &  13:53.46 &     28.47 \\
 6.059 & 17.062 &  13:55.39 &     28.02 \\
 5.801 & 16.939 &  13:56.25 &     27.9 \\
 5.824 & 20.069 &  13:57.32 &     27.66 \\
\bottomrule
\end{tabular}
\begin{tabular}{rrrr}
\toprule
     S/N MEAN & RMS MEAN & RMS STD \\
\midrule
    4.257 & 11.115 & 2.393 \\
\bottomrule
\end{tabular}
\\
\raggedright{
\textbf{Notes.} Residual images around the peak of the dynamic range have been manually inspected around the time of the potential transient and around any other time where data presented high dynamic range.}
\label{tab:tracks}
\end{table}

%% file: Appendix_plots.tex
\section{Snapshots of the transient search}
\label{sec:appendix_plots}

In this part of the Appendix, we show some of the snapshots with the highest S/N and rms. 
To avoid saturated frames and the loss of visual information, since the rms peak changes at each frame, we have normalized the color map scale to the peak of each frame.

\begin{figure*}
\centering
\includegraphics[scale=0.25]{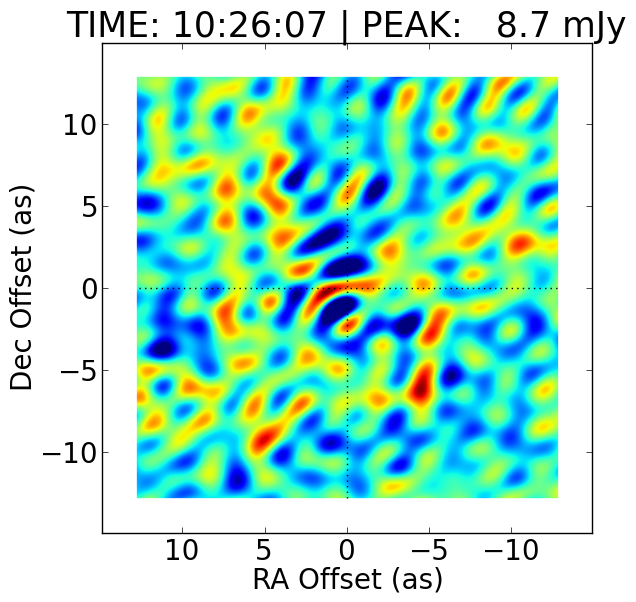}
\hspace{-0.2cm}
\includegraphics[scale=0.25]{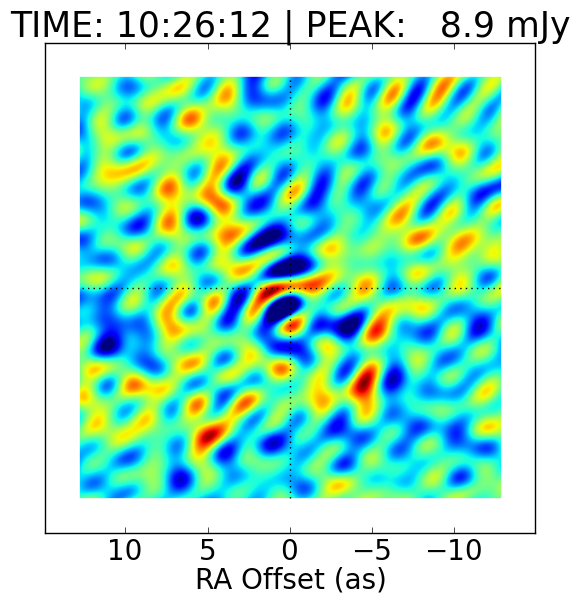}
\hspace{-0.2cm}
\includegraphics[scale=0.25]{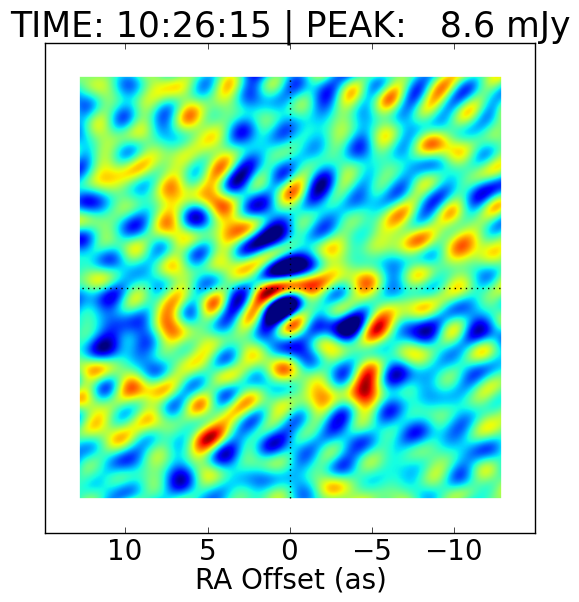}
\hspace{-0.2cm}
\includegraphics[scale=0.25]{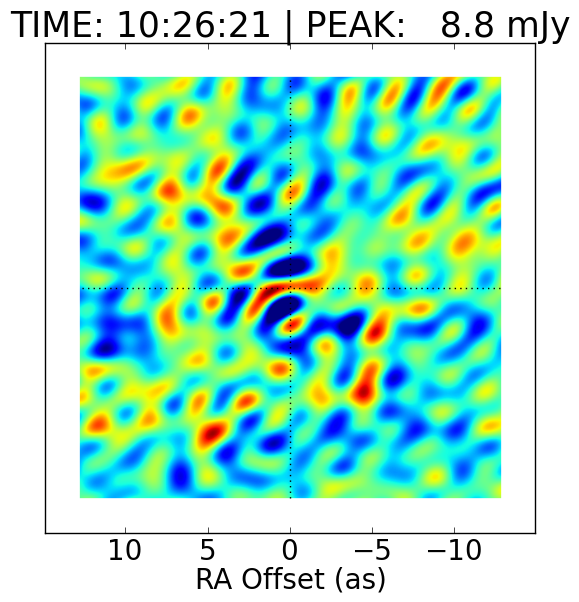}
\\
\includegraphics[scale=0.25]{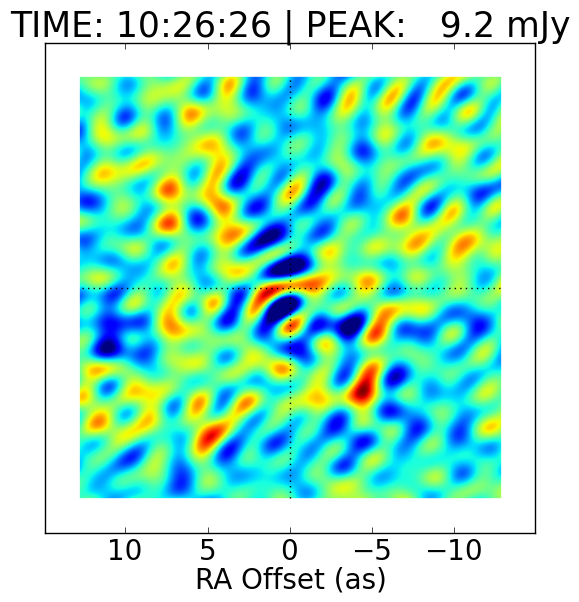}
\hspace{-0.2cm}
\includegraphics[scale=0.25]{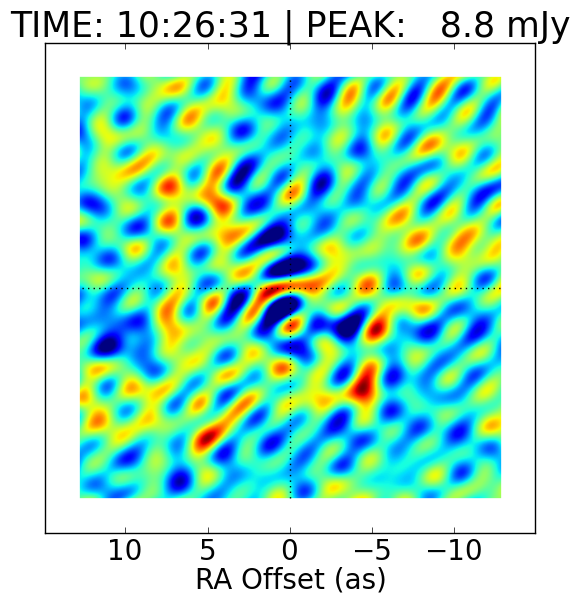}
\hspace{-0.2cm}
\includegraphics[scale=0.25]{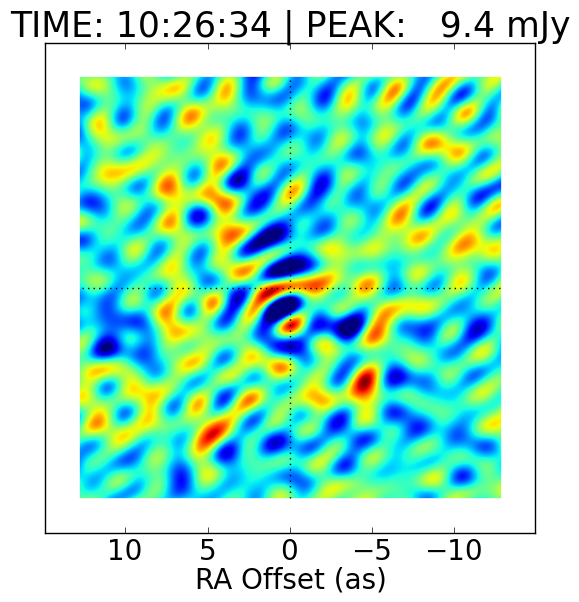}
\hspace{-0.2cm}
\includegraphics[scale=0.25]{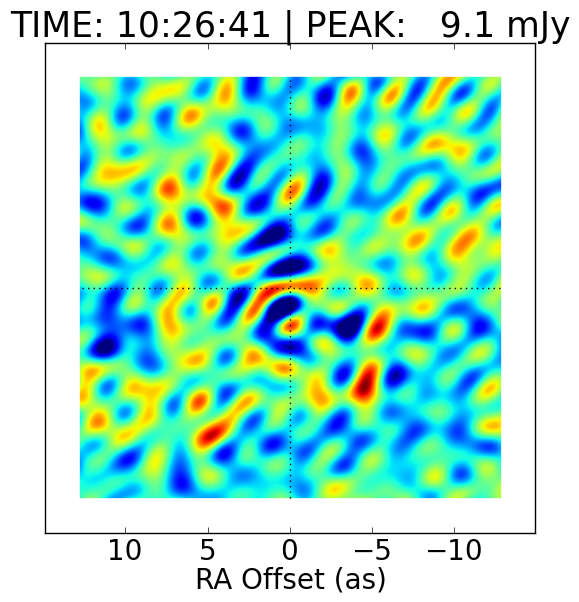}
\\
\includegraphics[scale=0.25]{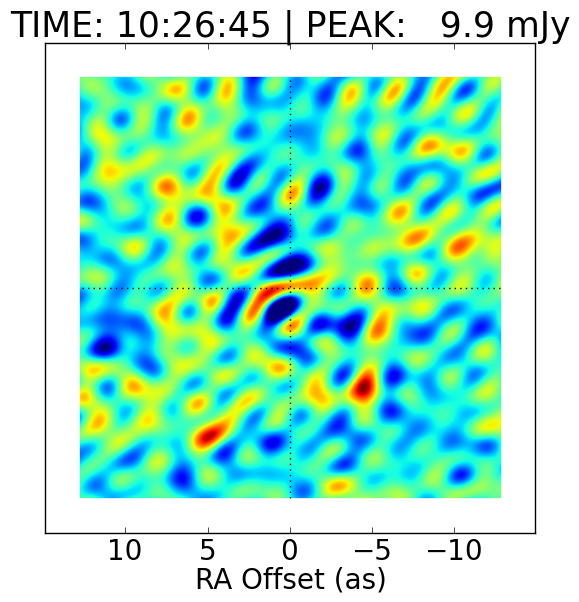}
\hspace{-0.2cm}
\includegraphics[scale=0.25]{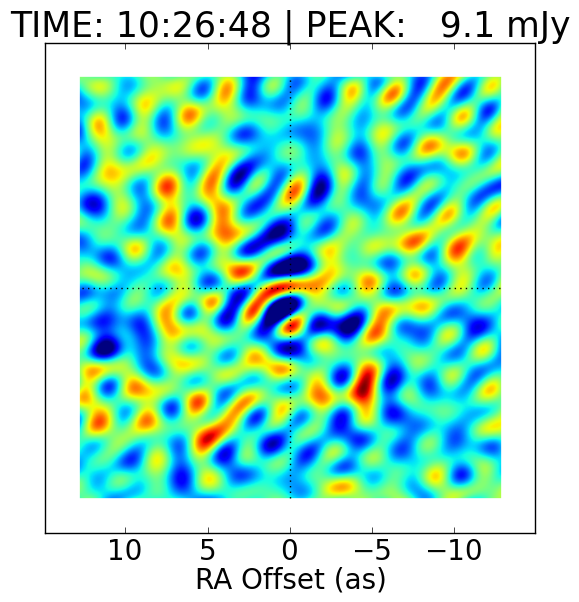}
\hspace{-0.2cm}
\includegraphics[scale=0.25]{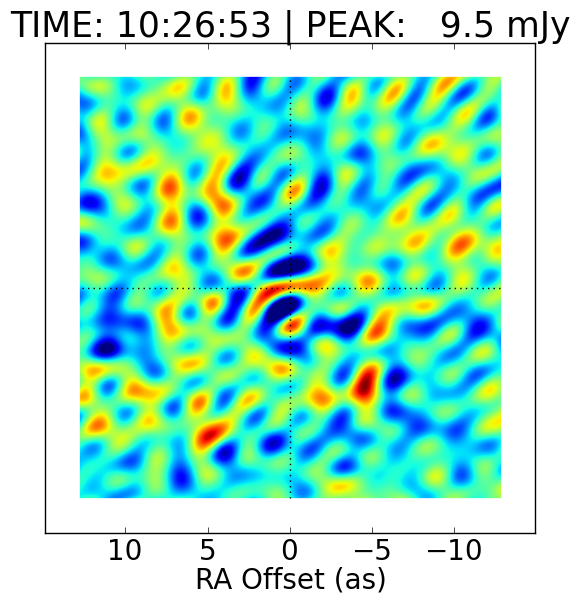}
\hspace{-0.2cm}
\includegraphics[scale=0.25]{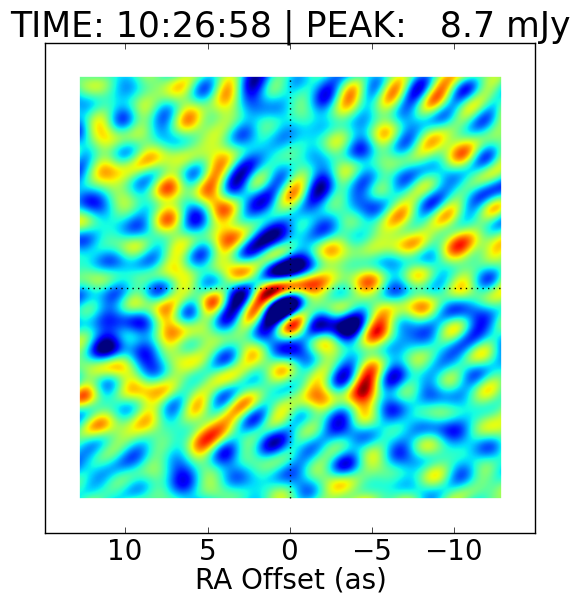}
\\
\includegraphics[scale=0.25]{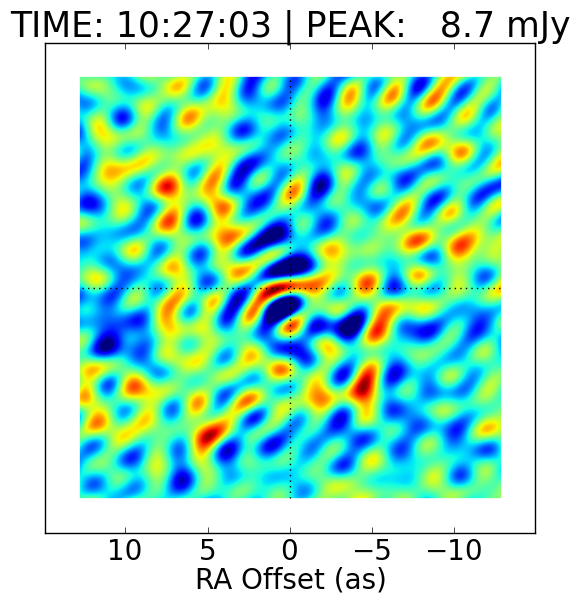}
\hspace{-0.2cm}
\includegraphics[scale=0.25]{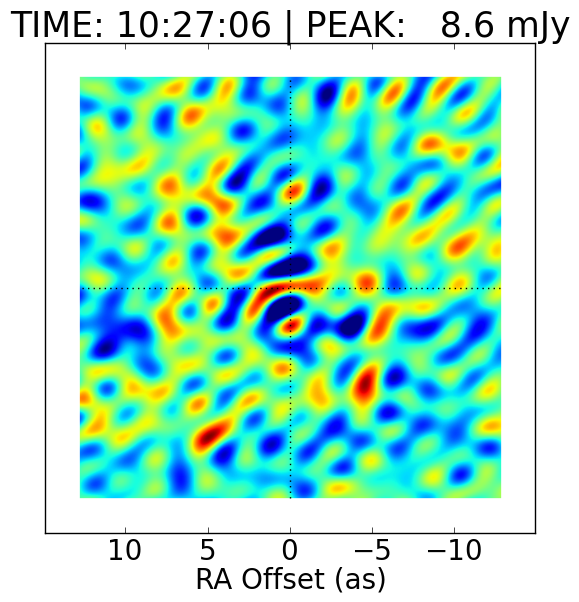}
\hspace{-0.2cm}

\hspace{-0.2cm}

\caption{Residuals of the frames of the track B when there could be misleading peaks in the flux.}
\label{fig:resb_fames}
\end{figure*}

Figure~\ref{fig:resb_fames} shows the 14 snapshots (with a cadence of 4\,s), of track B from 10:26:97 UT to 10:27:06 UT, where there was a maximum flux peak of 9.9\,mJy. As we can observe, none of the frames shows a clear isolated point-like source (which would indicate a transient origin), but rather a systematic (i.e., non-Gaussian) noise distribution (which indicates a dynamic range limitation, likely due to a sub-optimal gain calibration).

\begin{figure*}
\centering
\includegraphics[scale=0.25]{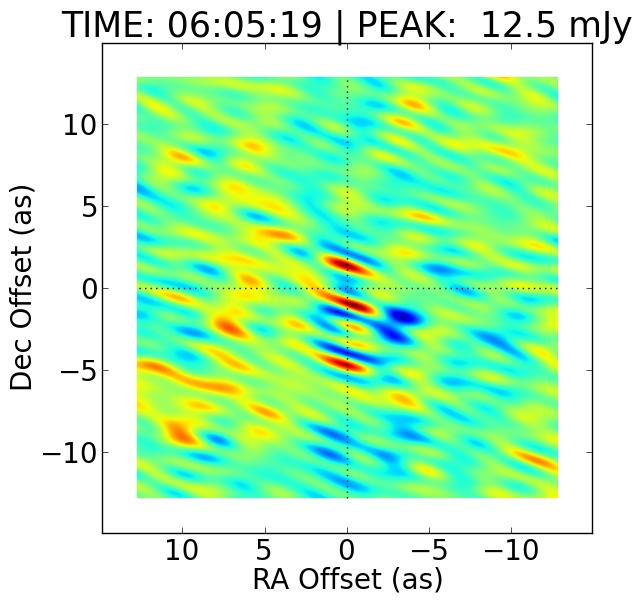}
\hspace{-0.2cm}
\includegraphics[scale=0.25]{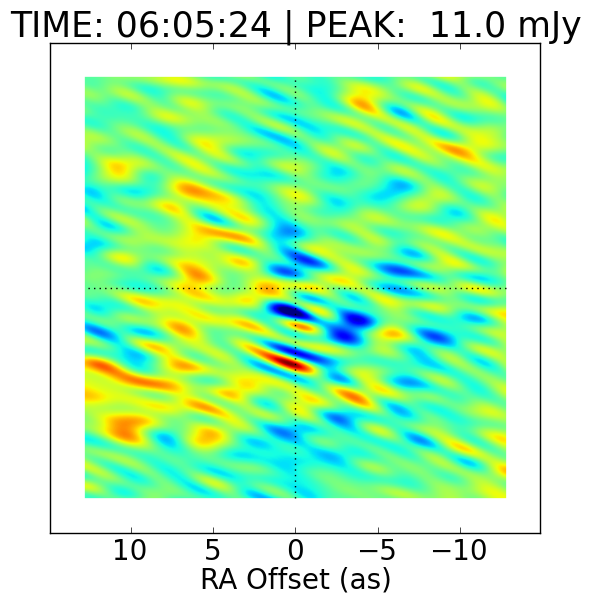}
\hspace{-0.2cm}
\includegraphics[scale=0.25]{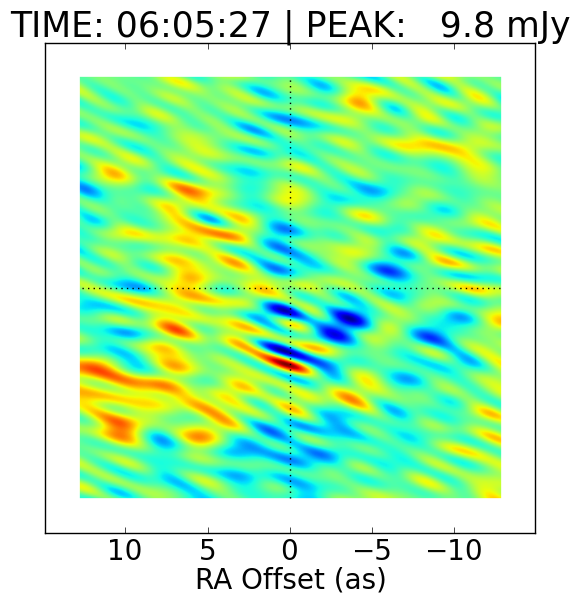}
\hspace{-0.2cm}
\includegraphics[scale=0.25]{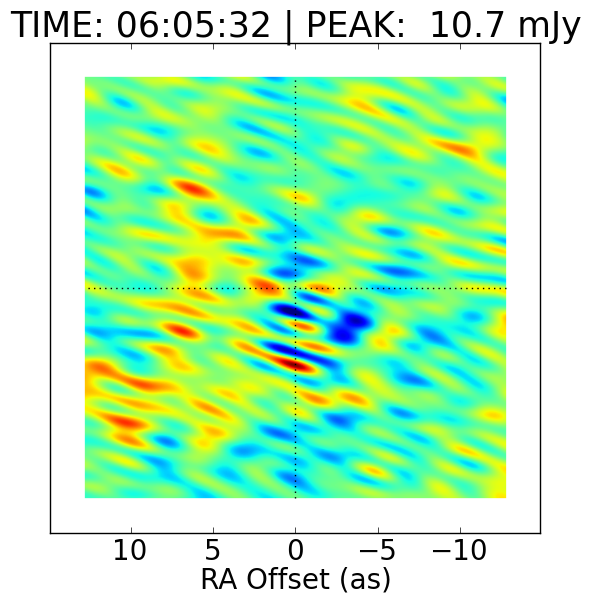}
\\
\includegraphics[scale=0.25]{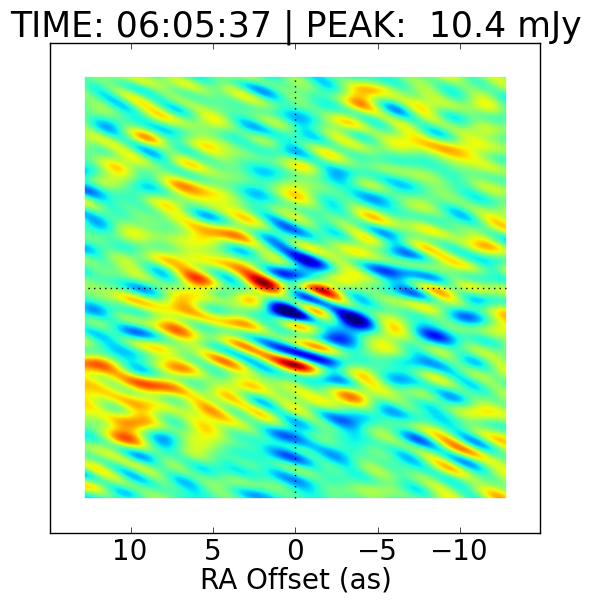}
\hspace{-0.2cm}
\includegraphics[scale=0.25]{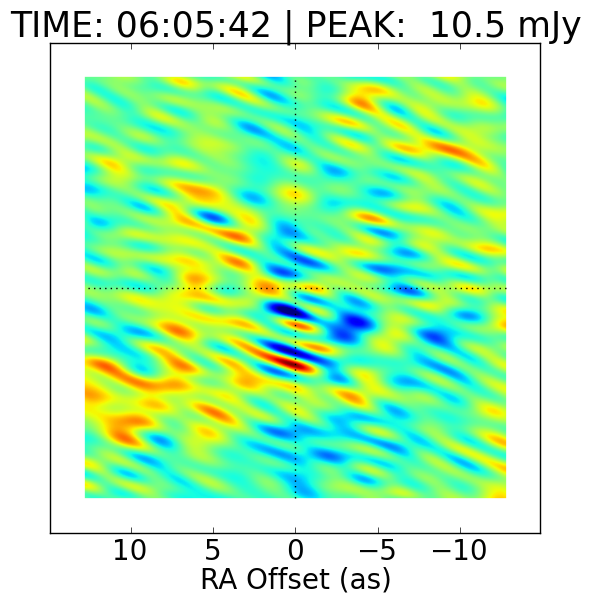}
\hspace{-0.2cm}
\includegraphics[scale=0.25]{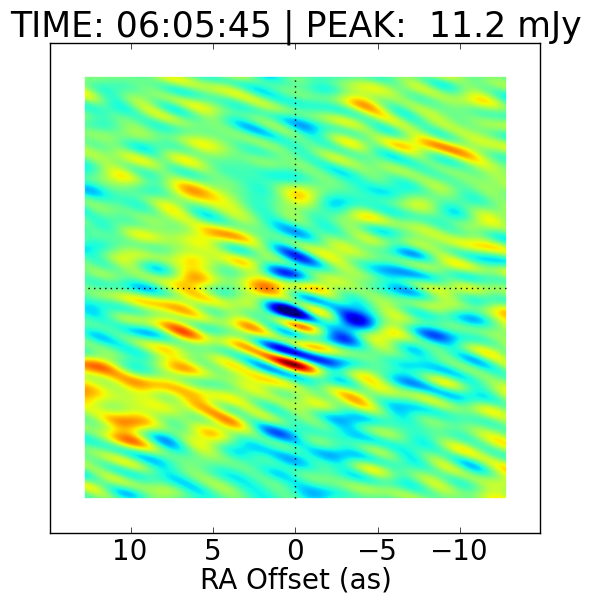}
\hspace{-0.2cm}
\includegraphics[scale=0.25]{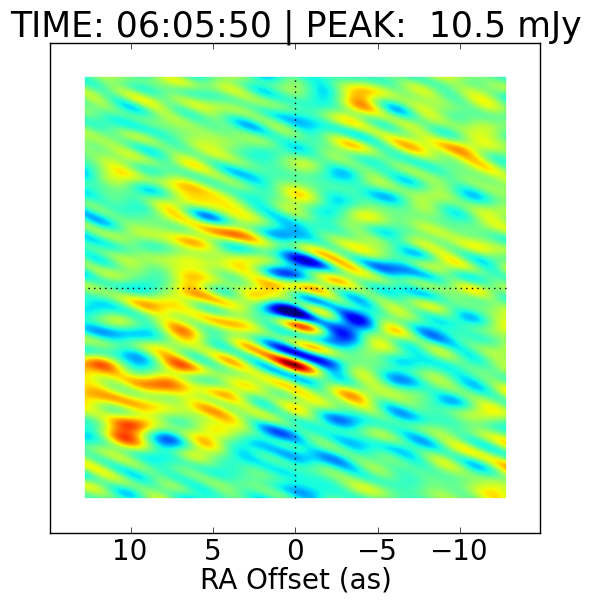}
\\
\includegraphics[scale=0.25]{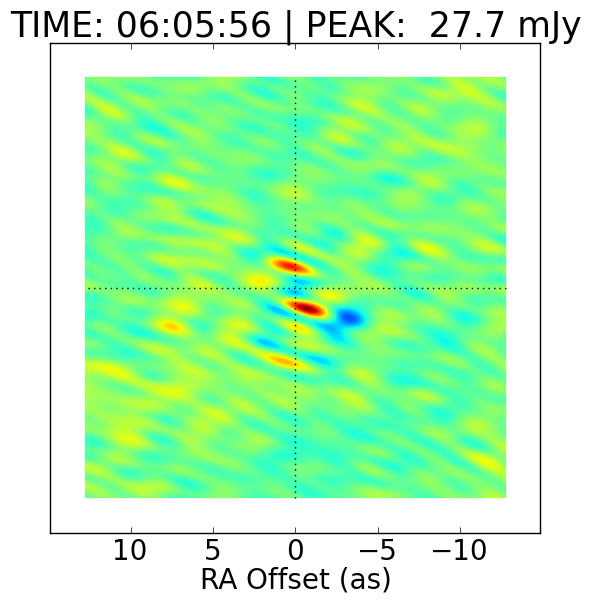}
\hspace{-0.2cm}
\includegraphics[scale=0.25]{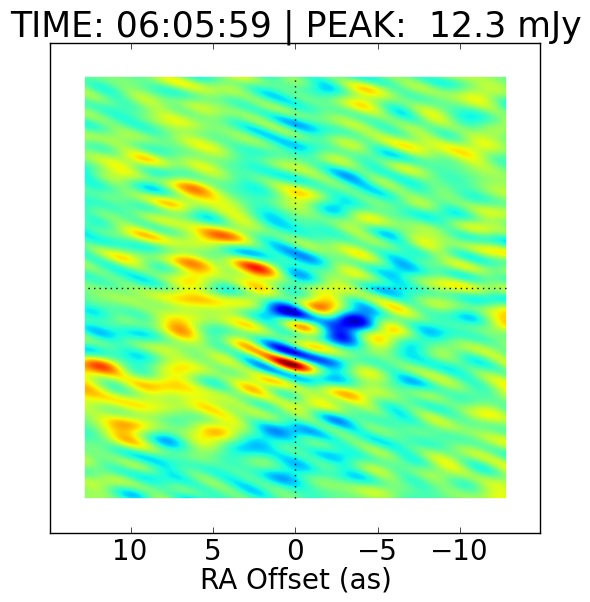}
\hspace{-0.2cm}
\includegraphics[scale=0.25]{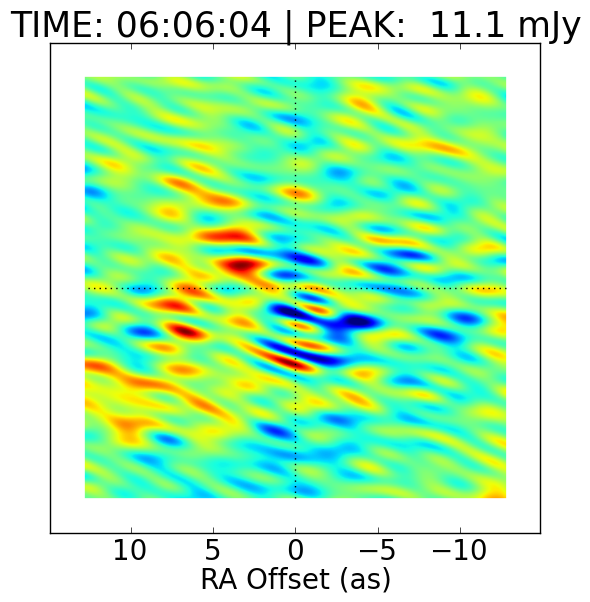}
\hspace{-0.2cm}
\includegraphics[scale=0.25]{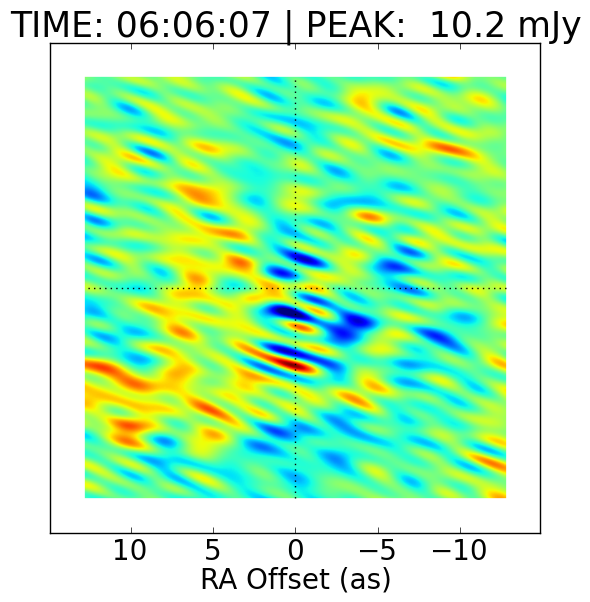}
\\
\includegraphics[scale=0.25]{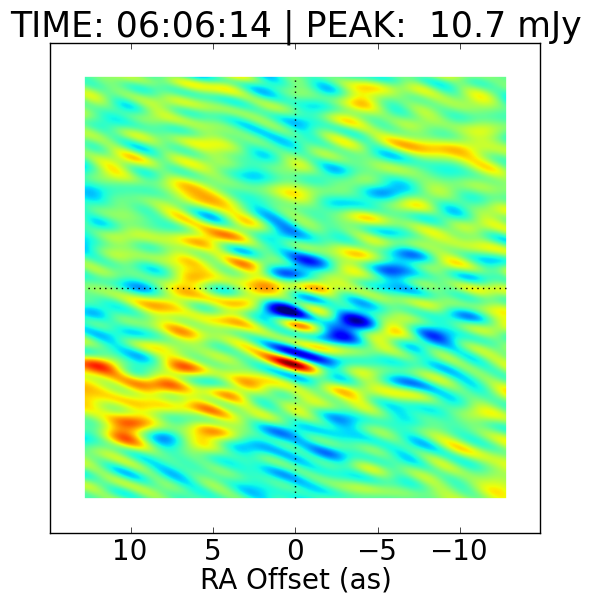}
\hspace{-0.2cm}
\includegraphics[scale=0.25]{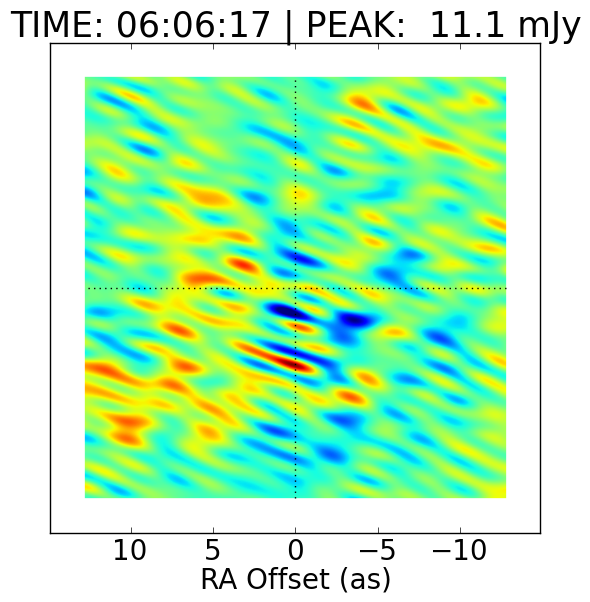}
\hspace{-0.2cm}

\caption{Residuals of the frames of the track C when there could be misleading peaks in the flux.}
\label{fig:resc_fames}
\end{figure*}

Figure~\ref{fig:resc_fames} are 14 snapshots taking from 06:15:19 h to 06:06:17 h. We notice a ``double source'' symmetrically distributed around the phase center (i.e., the location of SgrA*). This double-peak residual is likely due to a small (level of a 20$-$30 mJy) bias in the estimate of the SgrA* flux density at that observing time. In any case, this artifact only appears in one integration time, so it is discarded due to our duration constrain.

\begin{figure*}
\centering
\includegraphics[scale=0.25]{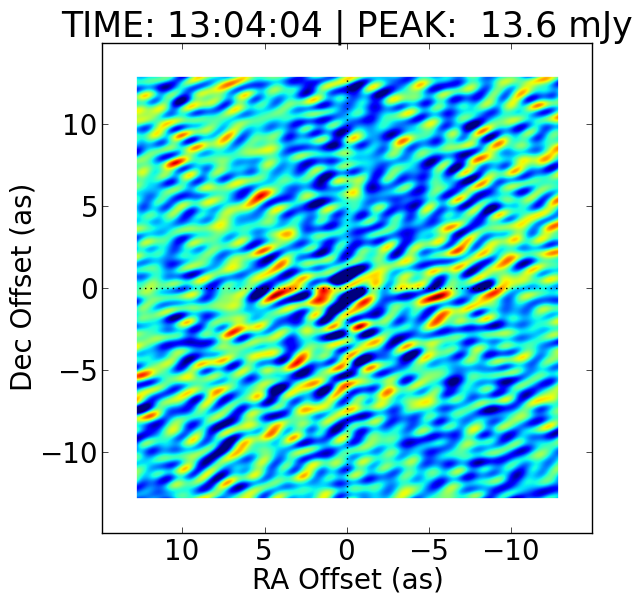}
\hspace{-0.2cm}
\includegraphics[scale=0.25]{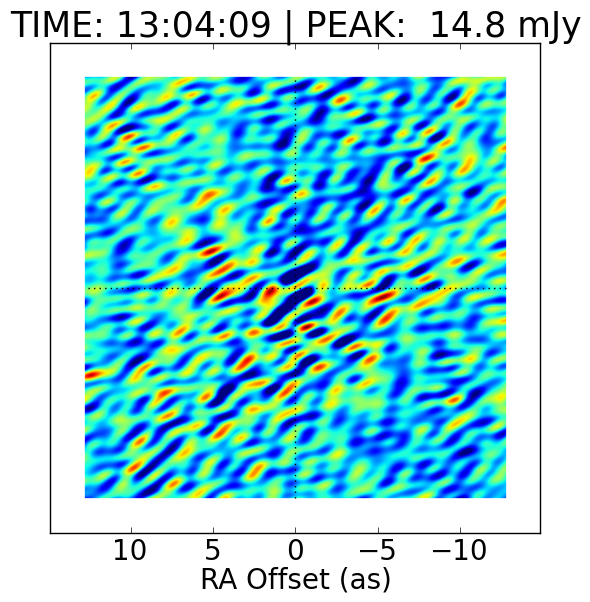}
\hspace{-0.2cm}
\includegraphics[scale=0.25]{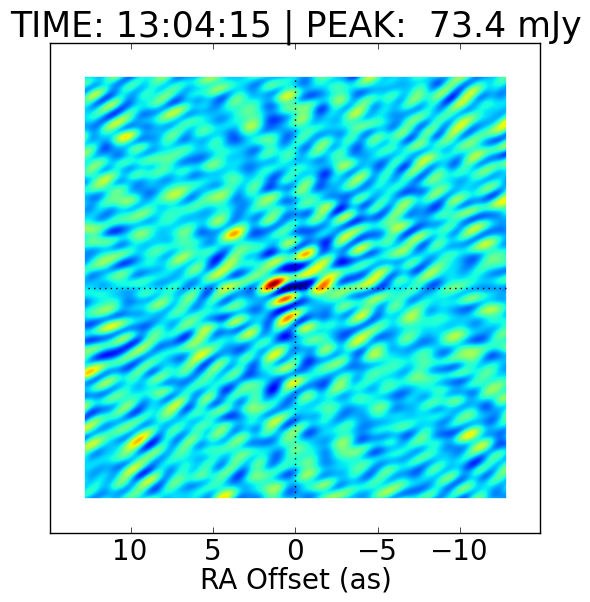}
\hspace{-0.2cm}
\includegraphics[scale=0.25]{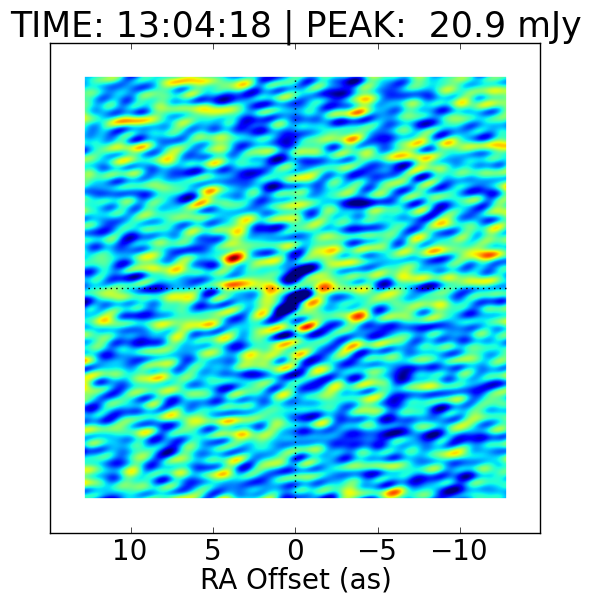}
\\
\includegraphics[scale=0.25]{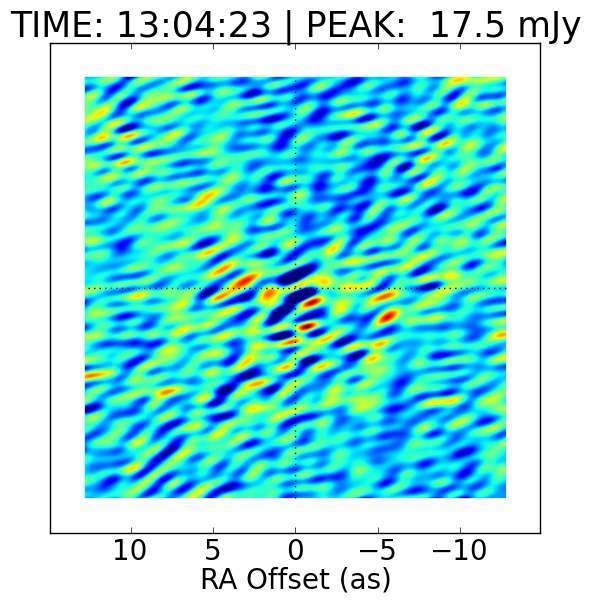}
\hspace{-0.2cm}
\includegraphics[scale=0.25]{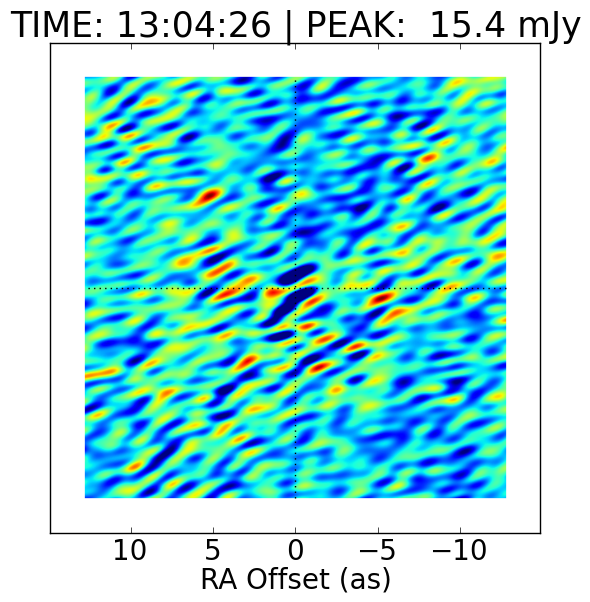}
\hspace{-0.2cm}
\includegraphics[scale=0.25]{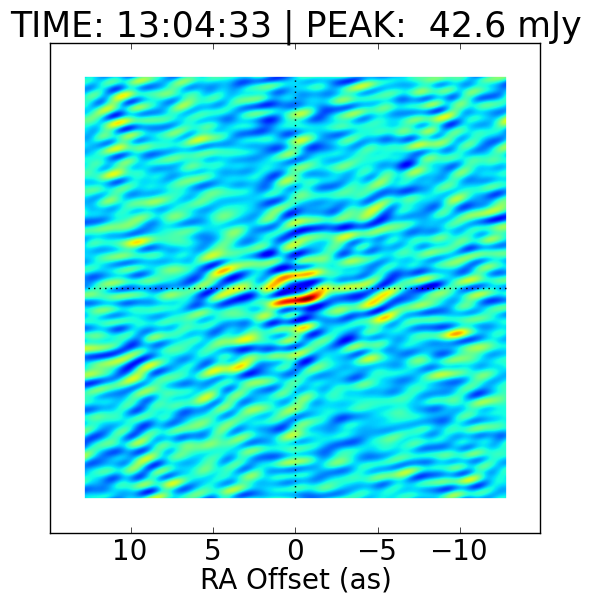}
\hspace{-0.2cm}
\includegraphics[scale=0.25]{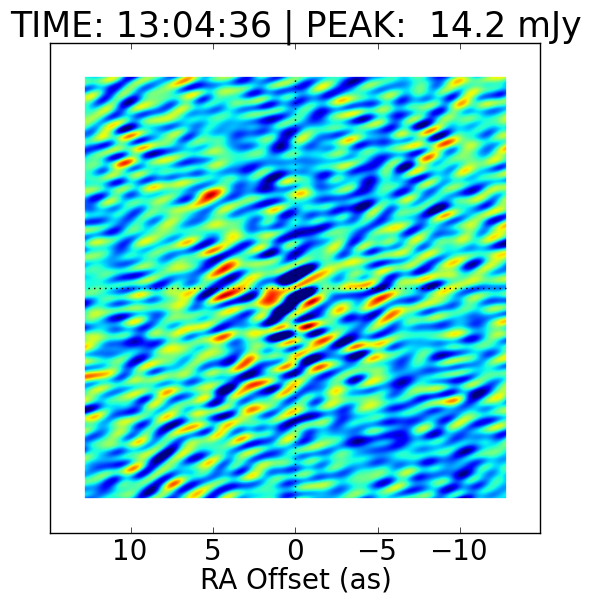}
\\
\includegraphics[scale=0.25]{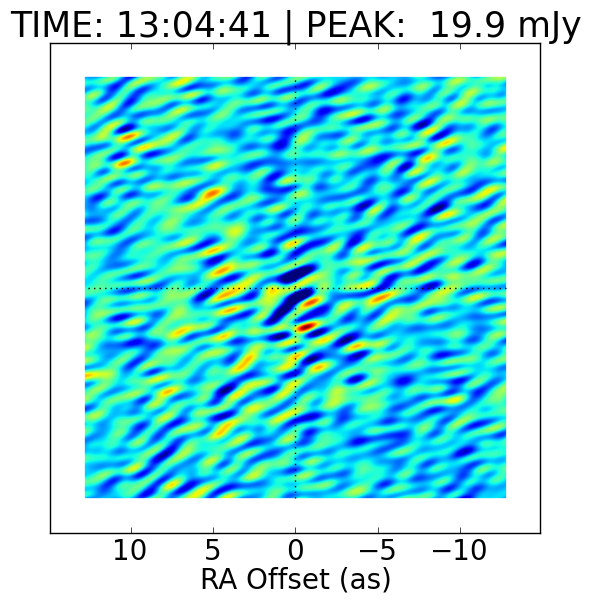}
\hspace{-0.2cm}
\includegraphics[scale=0.25]{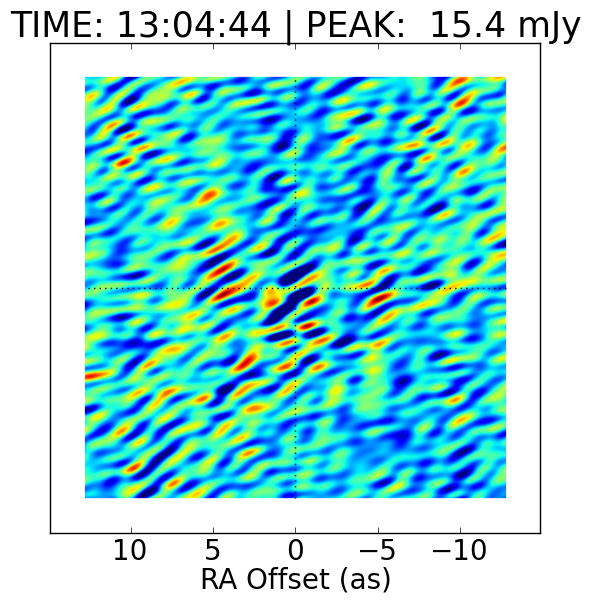}
\hspace{-0.2cm}
\includegraphics[scale=0.25]{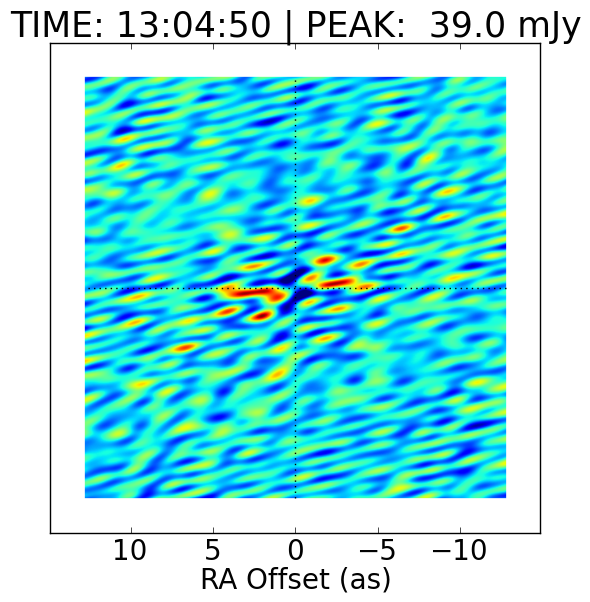}
\hspace{-0.2cm}
\includegraphics[scale=0.25]{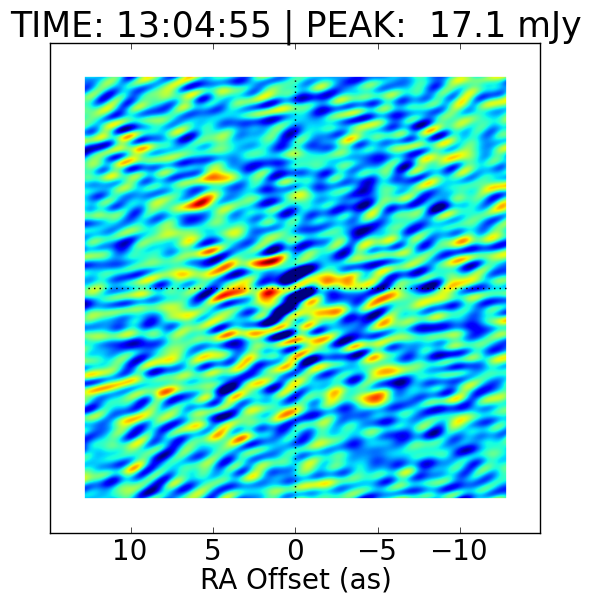}
\\
\includegraphics[scale=0.25]{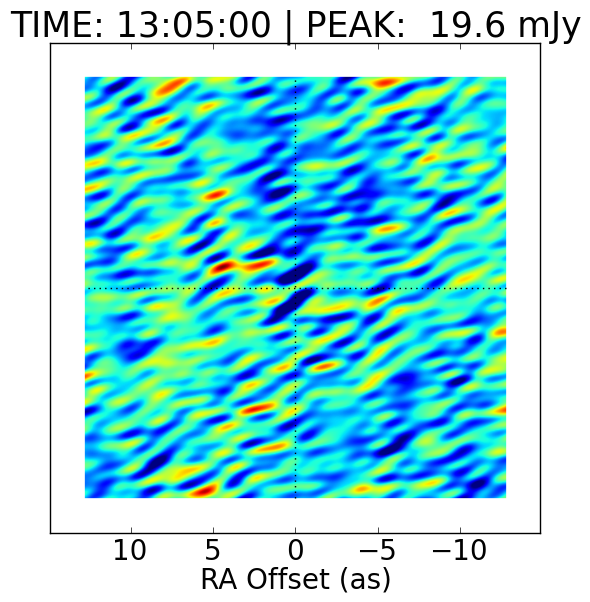}
\hspace{-0.2cm}
\includegraphics[scale=0.25]{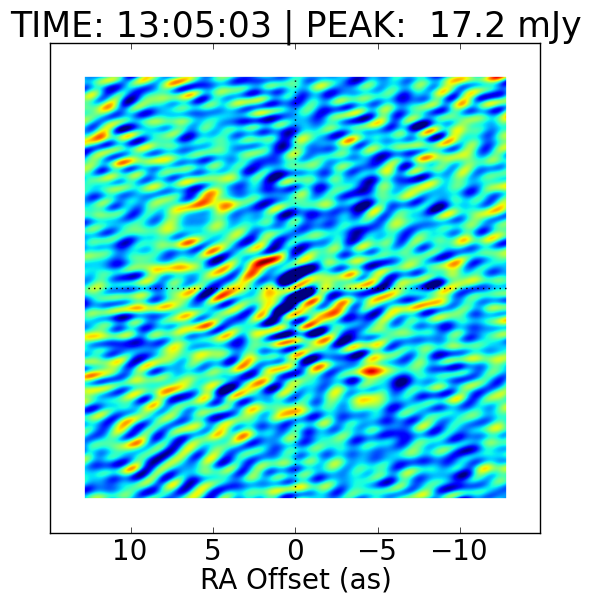}
\hspace{-0.2cm}

\caption{Residuals of the frames of the track E when there could be misleading peaks in the flux.}
\label{fig:rese_fames}
\end{figure*}

Finally, Figure~\ref{fig:rese_fames} shows again 14 snapshots starting from 13:04:09 h to 13:05:03 h of the last observation day, April 11 (track E). Recall that this day presented a particular higher noise conditions with respect to the other days. We can observe a higher mean of the peak in these frames with a maximum of 42\, mJy.